\documentclass{aa}
\usepackage{graphicx}
\usepackage{txfonts}
\usepackage{natbib}
\bibpunct{(}{)}{;}{a}{}{,}
\usepackage{longtable}
\usepackage{lscape}
\usepackage{rotating}

\begin{document}

\title{A Sino-German $\lambda$6\ cm polarization survey of the Galactic plane}
\subtitle{V. Large supernova remnants}
\author{X. Y.~Gao\inst{1}, J. L.~Han\inst{1}, W.~Reich\inst{2}, 
P.~Reich\inst{2}, X. H.~Sun\inst{1,2}, L.~Xiao\inst{1}}
\offprints{X. Y.~Gao: \email{bearwards@gmail.com}}

\institute{
National Astronomical Observatories, CAS, Jia-20 Datun Road, 
Chaoyang District, Beijing 100012, PR China 
\and
Max-Planck-Institut f\"{u}r Radioastronomie, 
Auf dem H\"{u}gel 69, 53121 Bonn, Germany
}

\date{Received; accepted}

\abstract 
{Observations of large supernova remnants (SNRs) at high frequencies
  are rare, but provide valuable information about their physical
  properties.}
{The total intensity and polarization properties of 16 large SNRs in
  the Galactic plane are investigated based on observations of the
  Urumqi $\lambda$6\ cm polarization survey of the Galactic plane with
  an angular resolution of 9$\farcm$5.}
{We extracted total intensity and linear polarization maps of large
  SNRs from the Urumqi $\lambda$6\ cm survey, obtained their
  integrated flux densities, and derived the radio spectra by using
  these flux densities together with previously published flux
  densities at various frequencies. In particular, Effelsberg
  $\lambda$11\ cm and $\lambda$21\ cm survey data were used to
  calculate integrated flux densities. The $\lambda$6\ cm polarization
  data also delineate the magnetic field structures of the SNRs.}
{We present the first total intensity maps at $\lambda$6\ cm for SNRs
  G106.3+2.7, G114.3+0.3, G116.5+1.1, G166.0+4.3 (VRO 42.05.01),
  G205.5+0.5 (Monoceros Nebula), and G206.9+2.3 (PKS 0646+06) and the
  first polarization measurements at $\lambda$6\ cm for SNRs G82.2+5.3
  (W63), G106.3+2.7, G114.3+0.3, G116.5+1.1, G166.0+4.3 (VRO
  42.05.01), G205.5+0.5 (Monoceros Nebula), and G206.9+2.3 (PKS
  0646+06). Most of the newly derived integrated radio spectra are
  consistent with previous results. The new flux densities obtained
  from the Urumqi $\lambda$6\ cm, Effelsberg $\lambda$11\ cm and
  $\lambda$21\ cm surveys are crucial to determine the spectra of SNR
  G65.1+0.6, G69.0+2.7 (CTB 80), G93.7-0.2, and G114.3+0.3. We find
  that G192.8$-$1.1 (PKS 0607+17) consists of background sources,
  \ion{H}{II} regions, and extended diffuse emission of thermal
  nature, and conclude that G192.8$-$1.1 is not a SNR.}
{}

\keywords {ISM: supernova remnants -- radio continuum: ISM -- 
Polarization}

\titlerunning{Urumqi $\lambda$6\ cm observations of large SNRs.}
\authorrunning{X. Y.~Gao et al.}

\maketitle
\section{Introduction}

A supernova explosion releases energy of about $10^{51}$~erg into the
interstellar space. Observations of supernova remnants (SNRs) are
important not only to the understanding of the objects themselves but
also to the properties of the ambient interstellar medium. Many intense
and large SNRs were identified in the first radio surveys at low
frequencies \citep[e.g.][]{Brown53, Westerhout58}. Sensitive and high
angular resolution Galactic plane surveys subsequently revealed a
large number of faint or compact SNRs \citep[e.g.][]{Reich88c,
  Brogan06}. To date, 274 Galactic SNRs have been catalogued by
\citet{Green09}. Observations of SNRs at many wavelengths are needed
to derive reliable spectra. Polarization observations of SNRs are
useful to reveal the magnetic field properties of SNRs.

The Sino-German $\lambda$6\ cm polarization survey of the Galactic
plane has been carried out with the Urumqi 25-m radio telescope since
2004. It covers $10\degr \leq \ell \leq 230\degr$ in Galactic
longitude and $|b| \leq 5\degr$ in Galactic latitude \citep{Sun07,
  Gao10, Sun11a, Xiao11}. The Urumqi $\lambda$6\ cm polarization
system is suitable for observations of large and faint SNRs. The
surface brightness limit for the survey is $\rm \Sigma_{1GHz} = 3.9\;
\times10^{-23} W m^{-2} Hz^{-1} sr^{-1}$, below the faintest SNRs
known today. 
Using this system, we have studied the magnetic fields and the
spectral index distribution of SNR G74.0$-$8.5 \citep[Cygnus Loop,
][]{Sun06}, G126.2+1.6 and G127.1+0.5 \citep{Sun07}, G156.2+5.7
\citep{Xu07}, G180.0$-$1.7 \citep[S147, ][]{Xiao08}, G130.7+3.1
\citep[HB3, ][]{Shi08a}, and G65.3+5.7 \citep{Xiao09}.  In addition,
\citet{Foster06} disapproved G166.2+2.5 (OA 184) as being a SNR.

In this paper, we analyze and discuss 16 large SNRs in the survey
region of $10\degr \leq \ell \leq 230\degr$ with an apparent size
exceeding $1\degr$ in our map. Smaller SNRs will be discussed
elsewhere (Sun et al. in prep).  We will briefly describe the survey
and the SNR sample in Sect.~2. In Sect.~3, we present their total
intensity and polarization images, and discuss the properties of each
SNR in detail. A summary is given in Sect.~4.

\begin{figure*}[!htb]
\begin{center}
\resizebox{0.4\textwidth}{!}{\includegraphics[angle=-90]{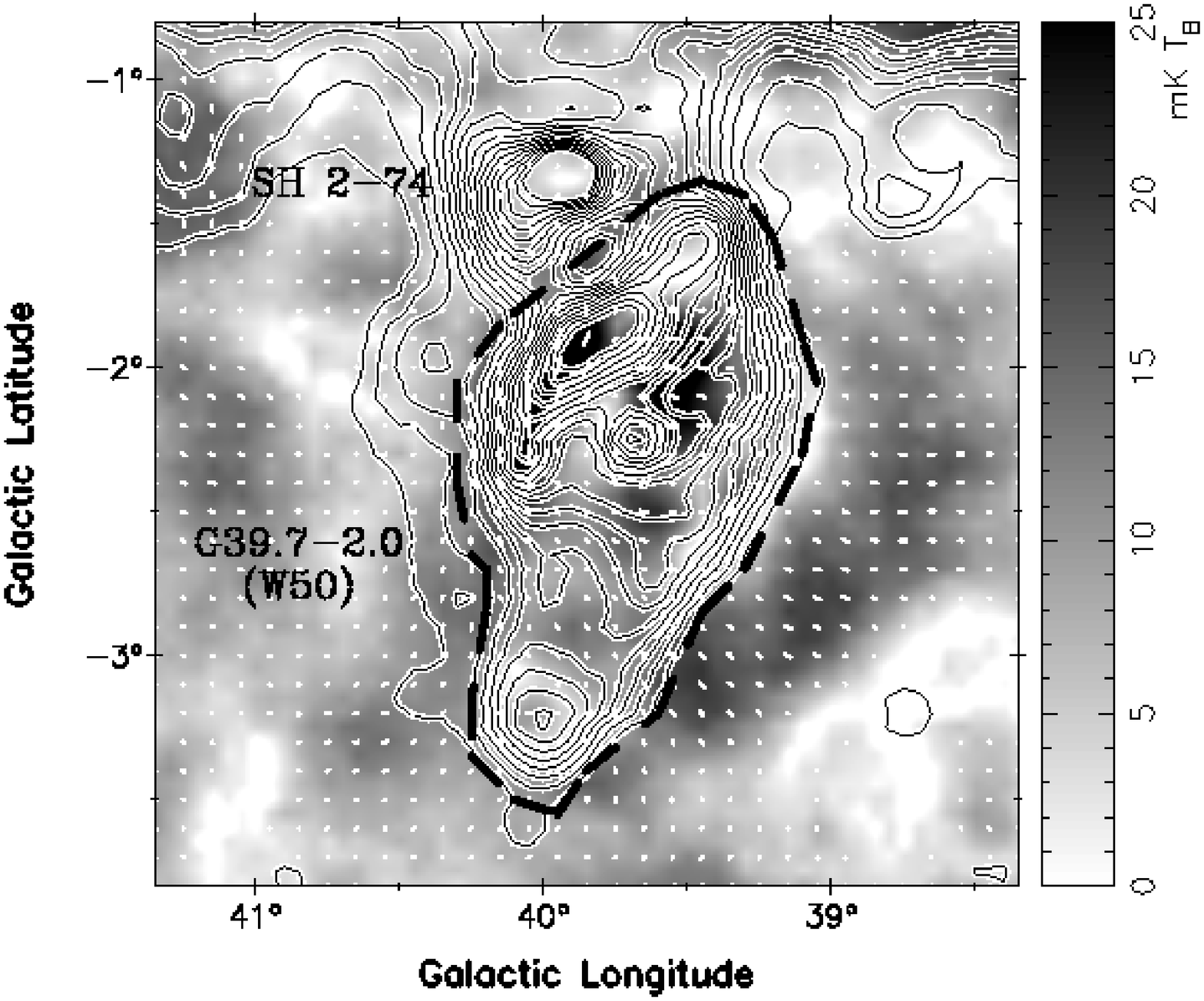}}
\resizebox{0.4\textwidth}{!}{\includegraphics[angle=-90]{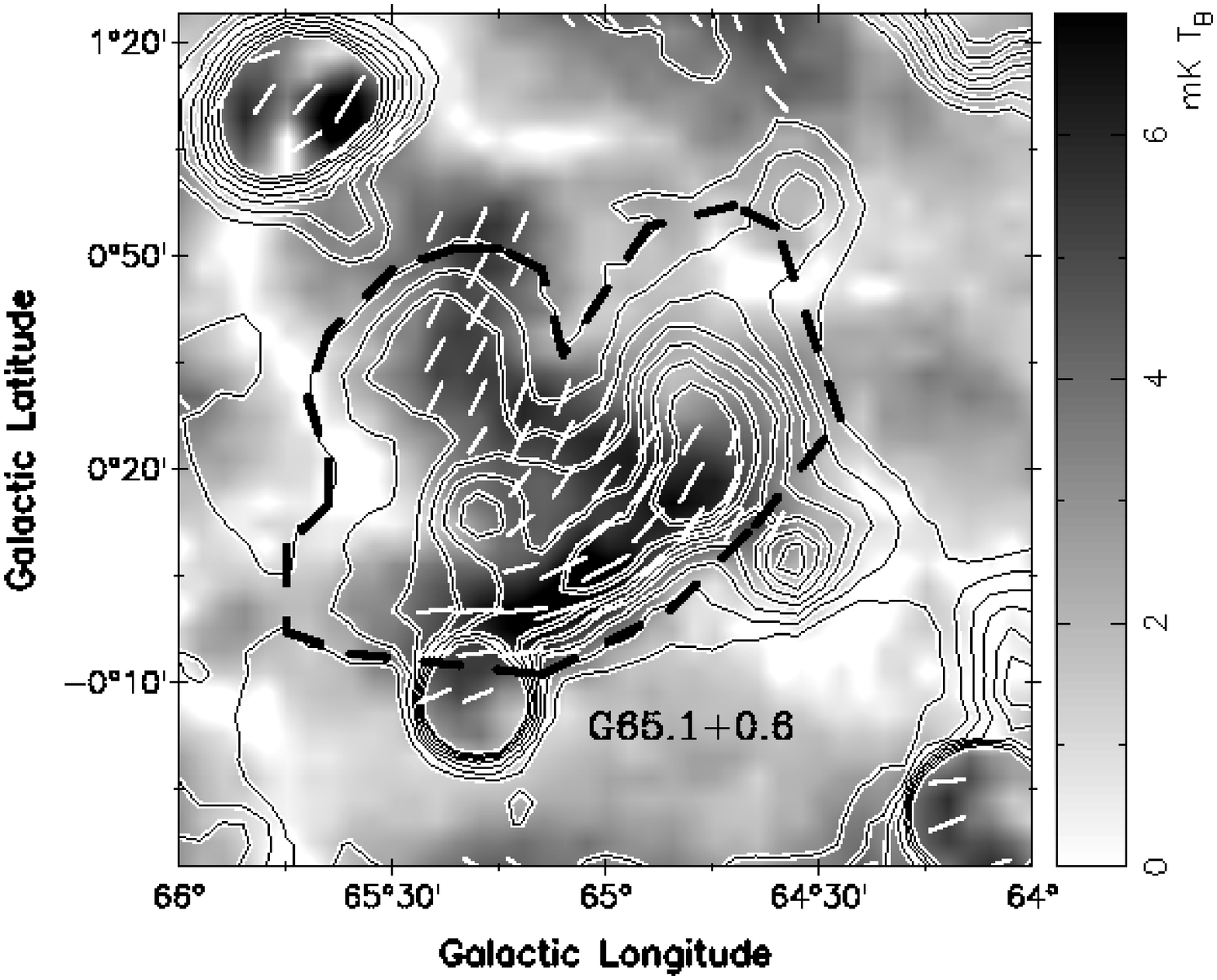}}\\
\resizebox{0.4\textwidth}{!}{\includegraphics[angle=-90]{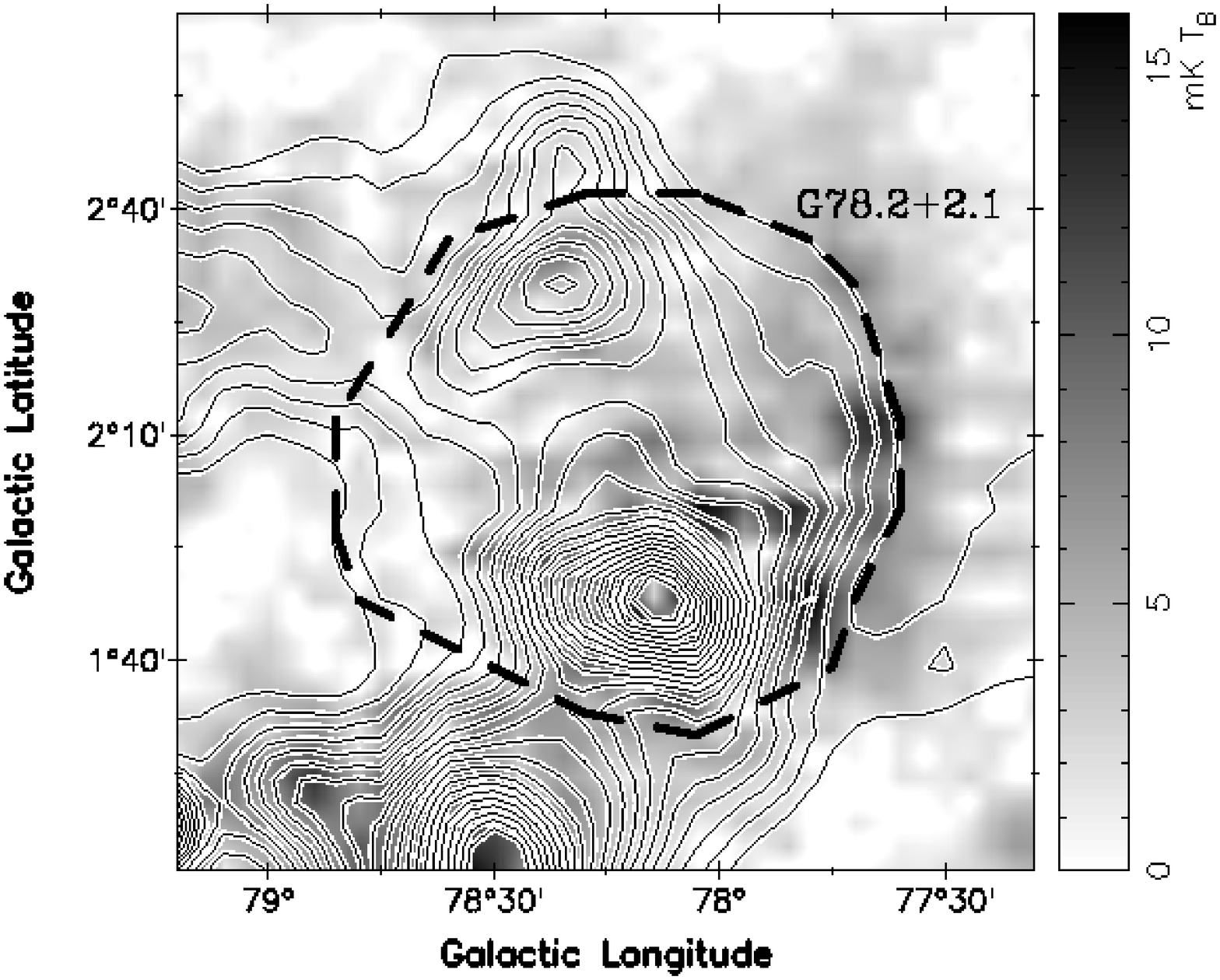}}
\resizebox{0.4\textwidth}{!}{\includegraphics[angle=-90]{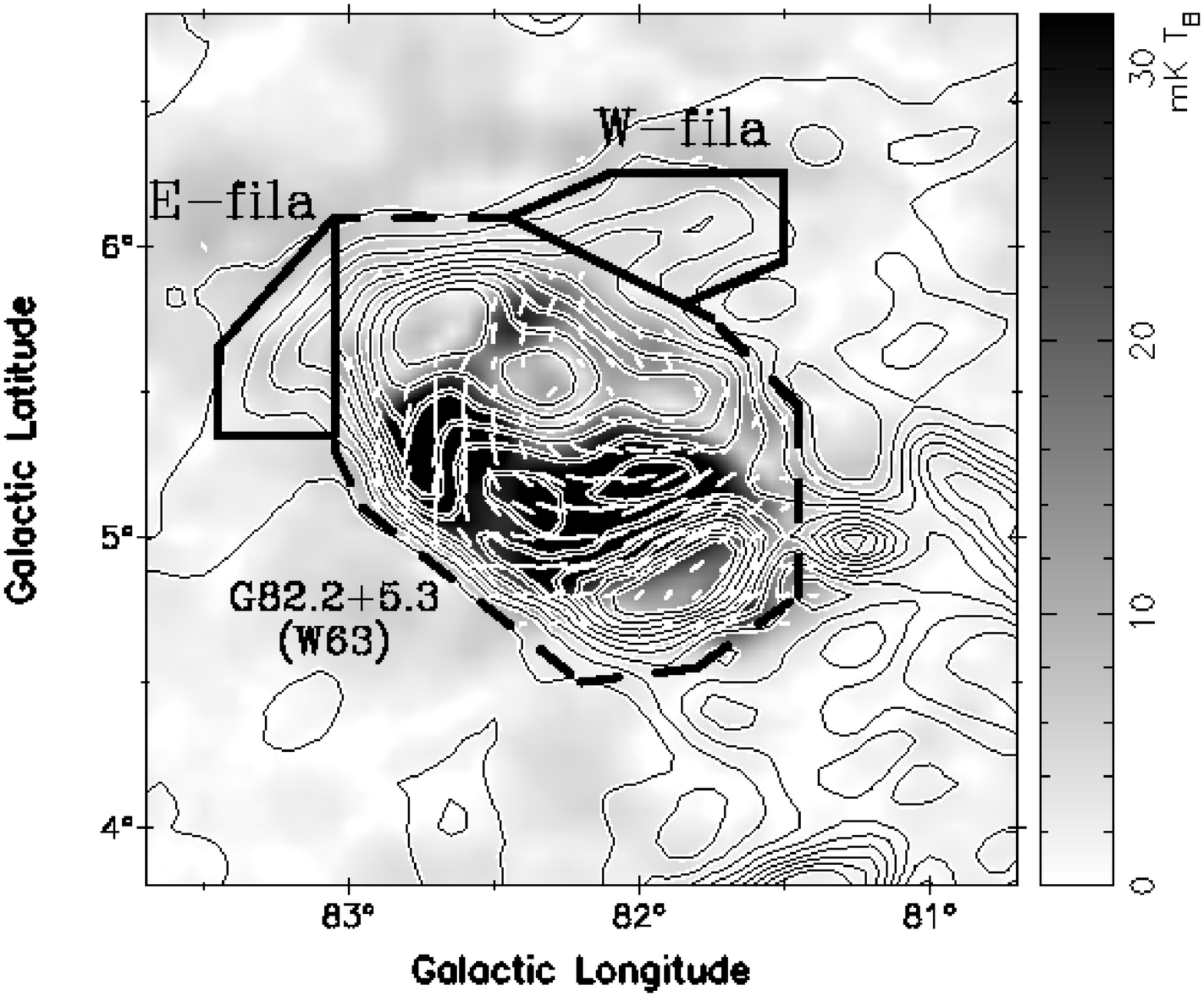}}\\
\resizebox{0.4\textwidth}{!}{\includegraphics[angle=-90]{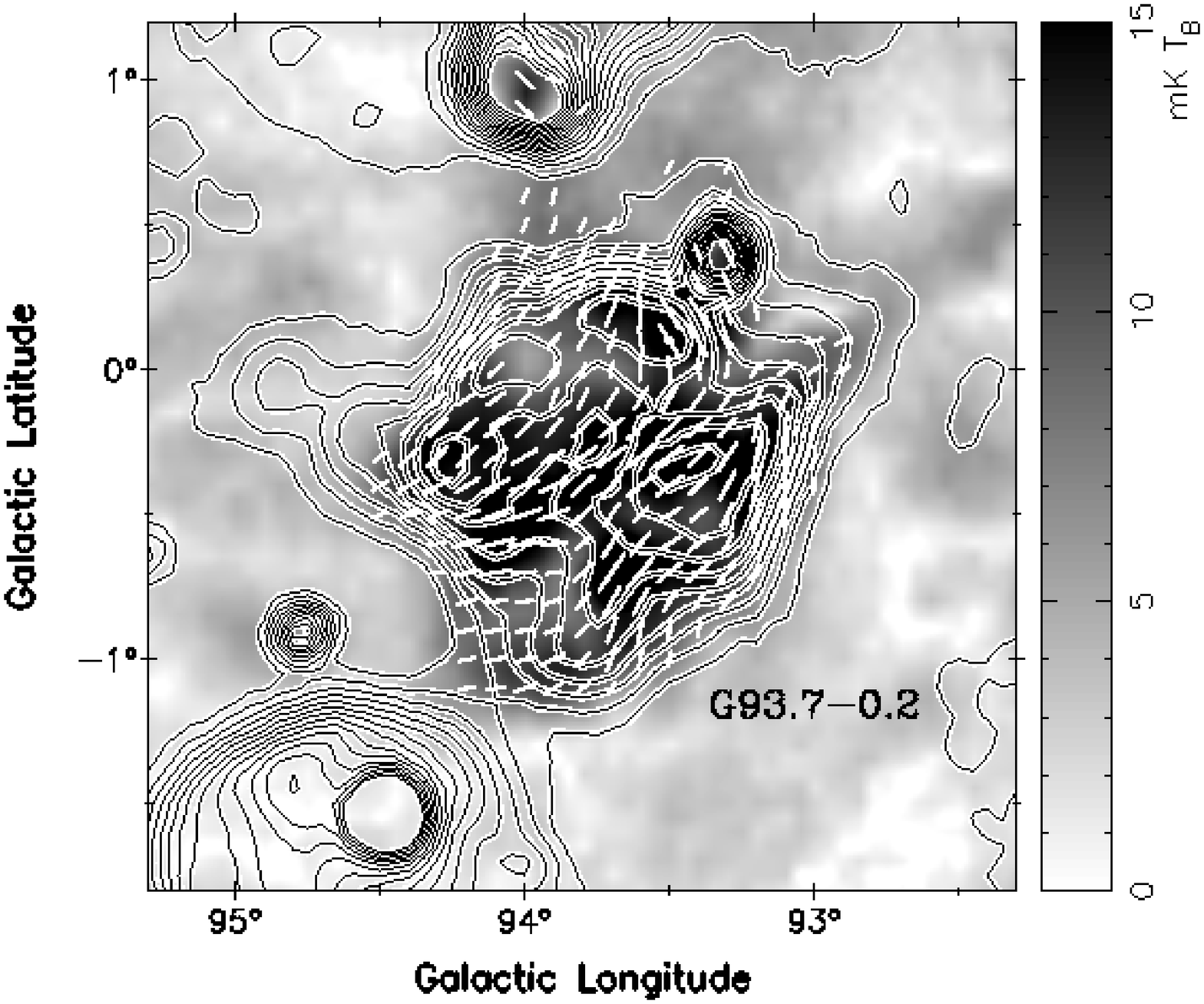}}
\resizebox{0.4\textwidth}{!}{\includegraphics[angle=-90]{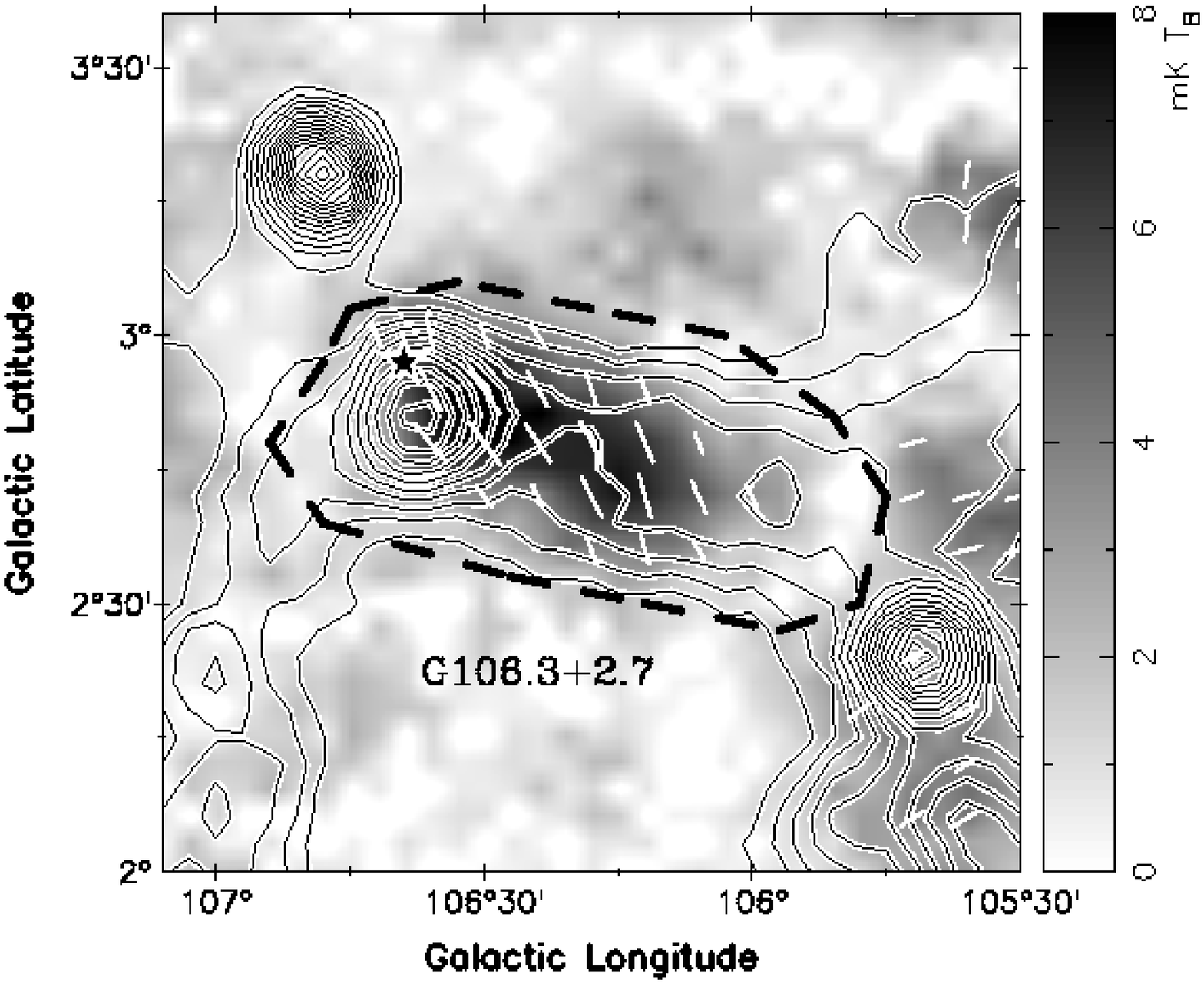}}\\
\end{center}
\caption{Polarization intensity (PI) in greyscale images overlaid by
  total intensity contours for 16 SNRs (more in next pages).  Four of
  them, G69.0+2.7 (CTB 80), G89.0+4.7 (HB21), G160.9+2.6 (HB9), and
  G205.5+0.5 (Monoceros Nebula), are presented in page 4 by slightly
  larger plots.  Observed polarization B-field vectors (i.e. $PA$ of
  the observed E-vector plus $90\degr$) of SNRs were overlaid every
  $6\arcmin$ in case PI exceeds a certain threshold (see below). Their
  lengths are proportional to PI. The stars indicate the pulsar in
  G114.3+0.3 and pulsar wind nebula inside G106.3+2.7, IC443 and CTB
  80.  The thick dashed line marks the boundary for flux density
  integration for G39.7$-$2.0 (W50), G65.1+0.6, G78.2+2.1, G82.2+5.3
  (W63), G106.3+2.7, G189.1+3.0 (IC443), CTB 80, and the western shell
  of the Monoceros Nebula. All contour lines run from the local
  $3\sigma$ level (see values for each SNR below) in different steps.
  The contour start level, contour step, and PI threshold for plotted
  B-vectors (all in $\rm mK\ T_{B}$) are for W50: 13.5, 13.5, 3.0; for
  G65.1+0.6: 3.9, 5.0, 4.0; for G78.2+2.1: 90.0, 120.0, 15.0; for W63:
  9.0, 24.0, 8.0; for G93.7$-$0.2: 4.5, 12.0, 7.0; for G106.3+2.7:
  6.0, 4.2, 3.2; for G114.3+0.3: 4.0, 4.0, 3.8; for G116.5+1.1: 3.6,
  4.5, 7.0; for G166.0+4.3 (VRO 42.05.01): 2.1, 5.0, 3.5; for
  G179.0+2.6: 1.5, 3.5, 2.5; for G206.9+2.3 (PKS 0646+06): 3.0, 4.0,
  3.0; for HB21: 3.3, 34.0, 7.5; for HB9: 5.0, 8.0, 7.0. For the
  Monoceros Nebula: 4.0, 6.0, 4.5. For IC443, contours start at $\rm
  7.2~mK\ T_{B}$ and run in steps of $\rm 2^{n-1}\times4.8~mK~T_{B}$
  (n = 1, 2, 3 $\ldots$), the PI threshold is $\rm 7.0~mK\ T_{B}$. For
  CTB 80, contours start at $\rm 3.3~mK\ T_{B}$ and run in steps of
  $\rm 2^{\frac{n-1}{2}}\times3.3~mK\ T_{B}$, and the PI threshold is
  $\rm 10.0~mK\ T_{B}$.}
\label{SNRipi}
\end{figure*}\addtocounter{figure}{-1}
\begin{figure*}
\begin{center}
\resizebox{0.46\textwidth}{!}{\includegraphics[angle=-90]{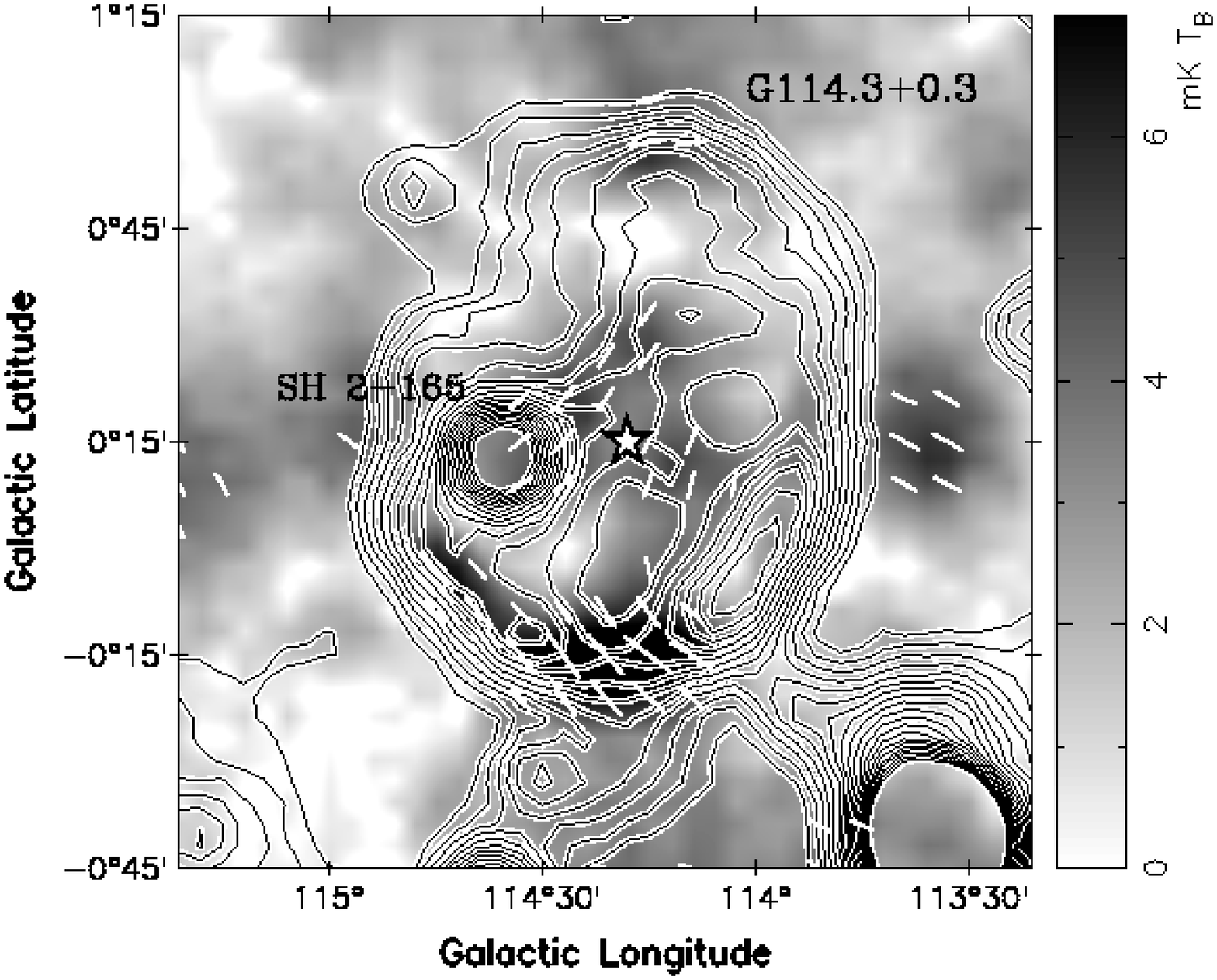}}
\resizebox{0.46\textwidth}{!}{\includegraphics[angle=-90]{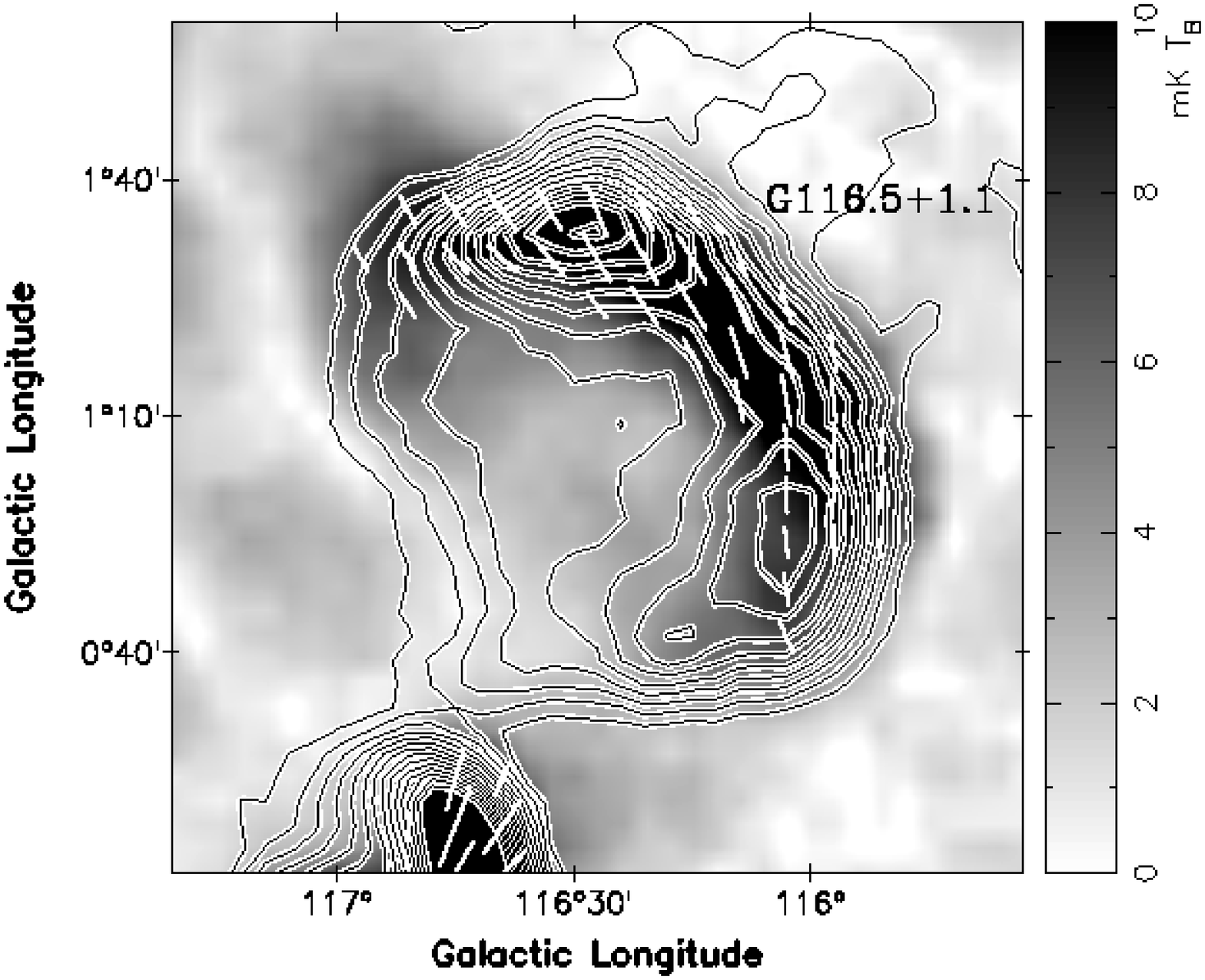}}\\
\resizebox{0.46\textwidth}{!}{\includegraphics[angle=-90]{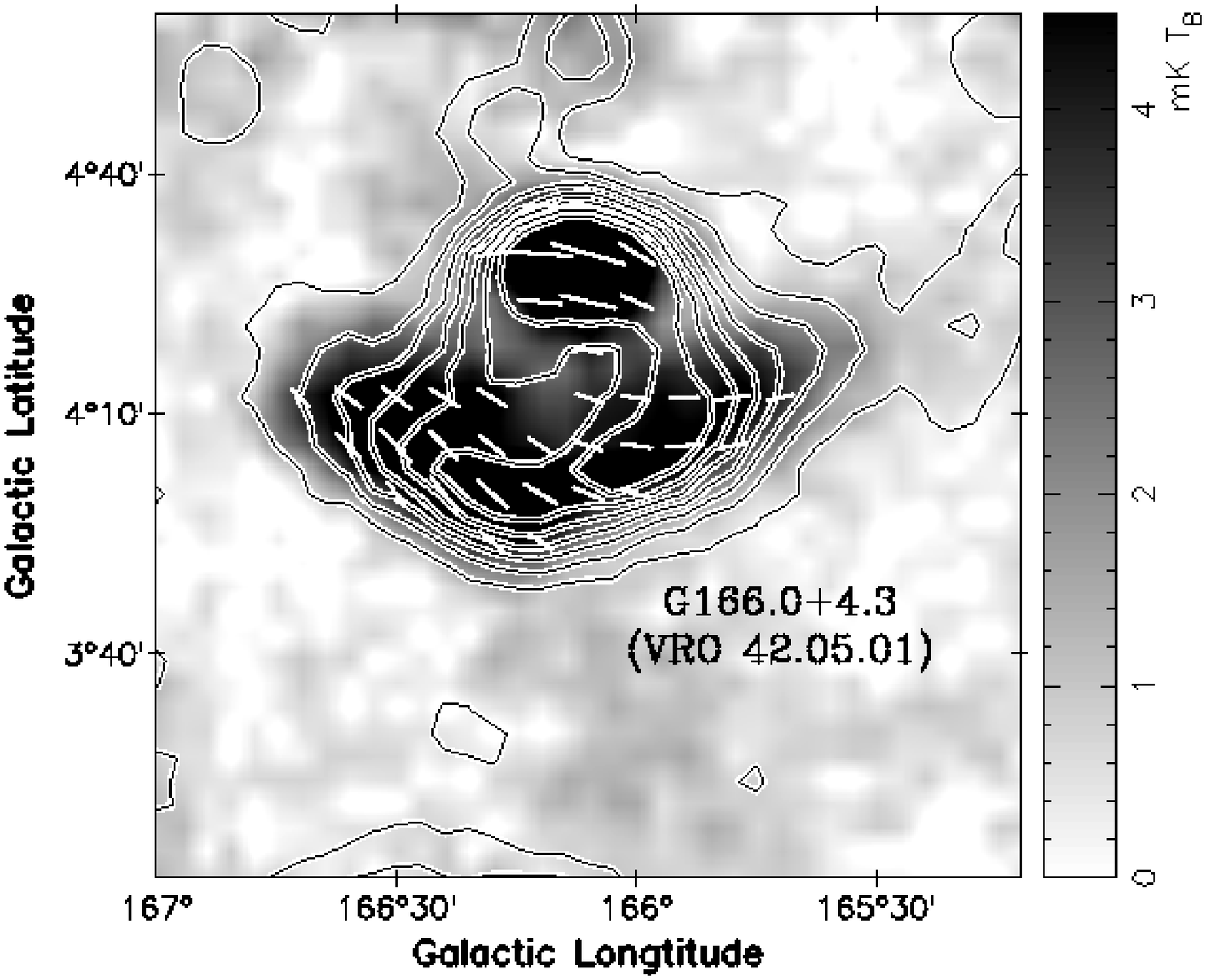}}
\resizebox{0.46\textwidth}{!}{\includegraphics[angle=-90]{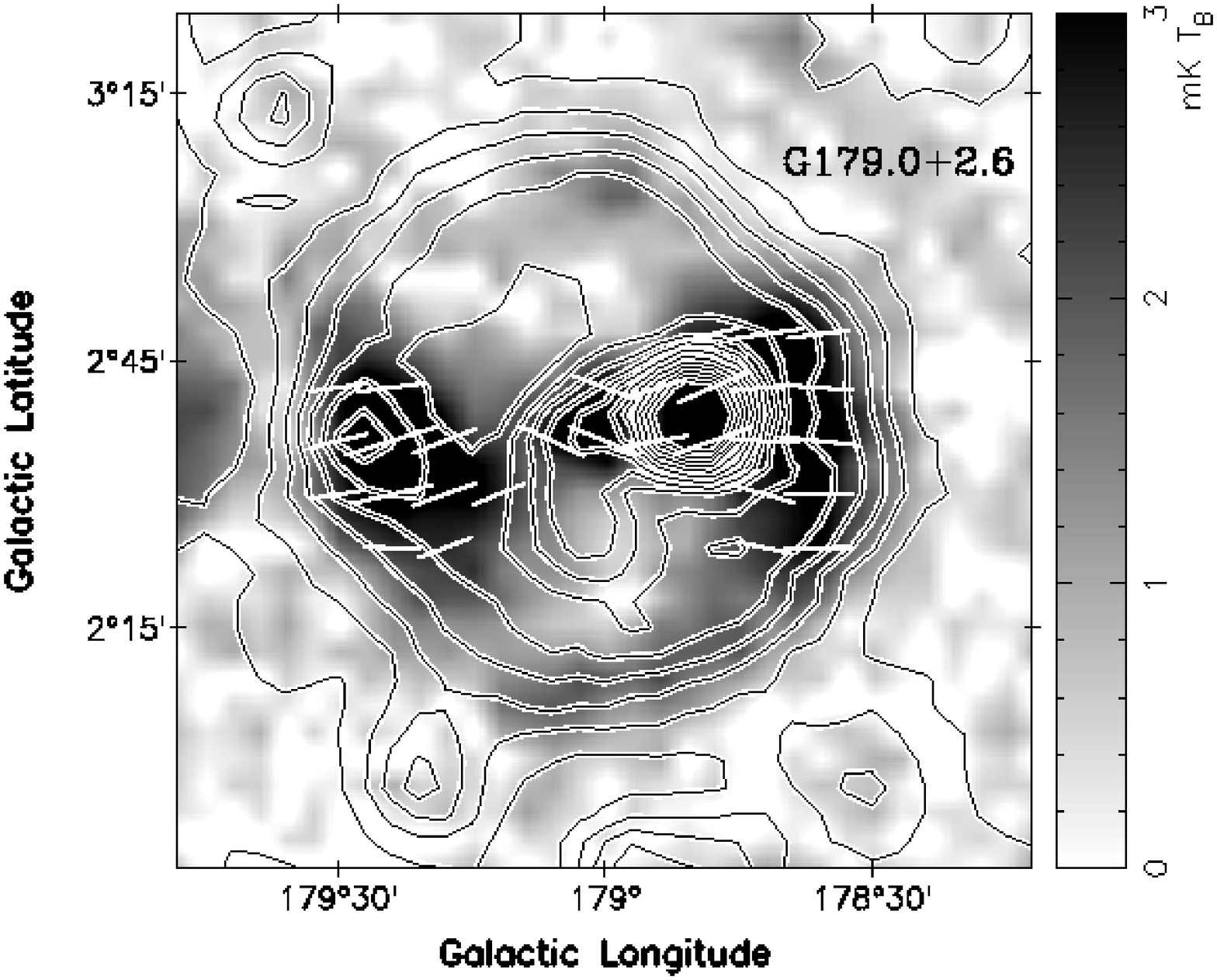}}\\
\resizebox{0.46\textwidth}{!}{\includegraphics[angle=-90]{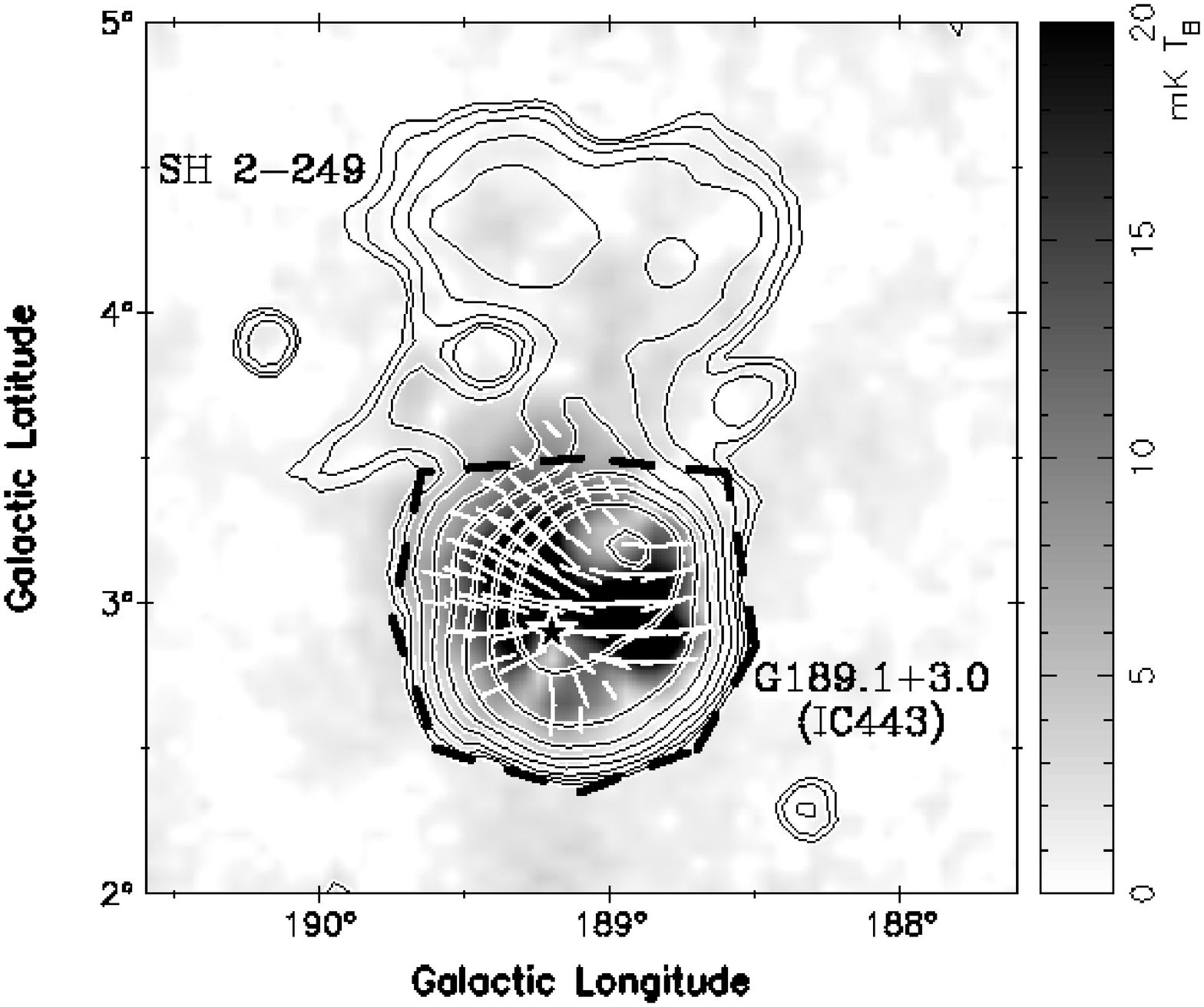}}
\resizebox{0.46\textwidth}{!}{\includegraphics[angle=-90]{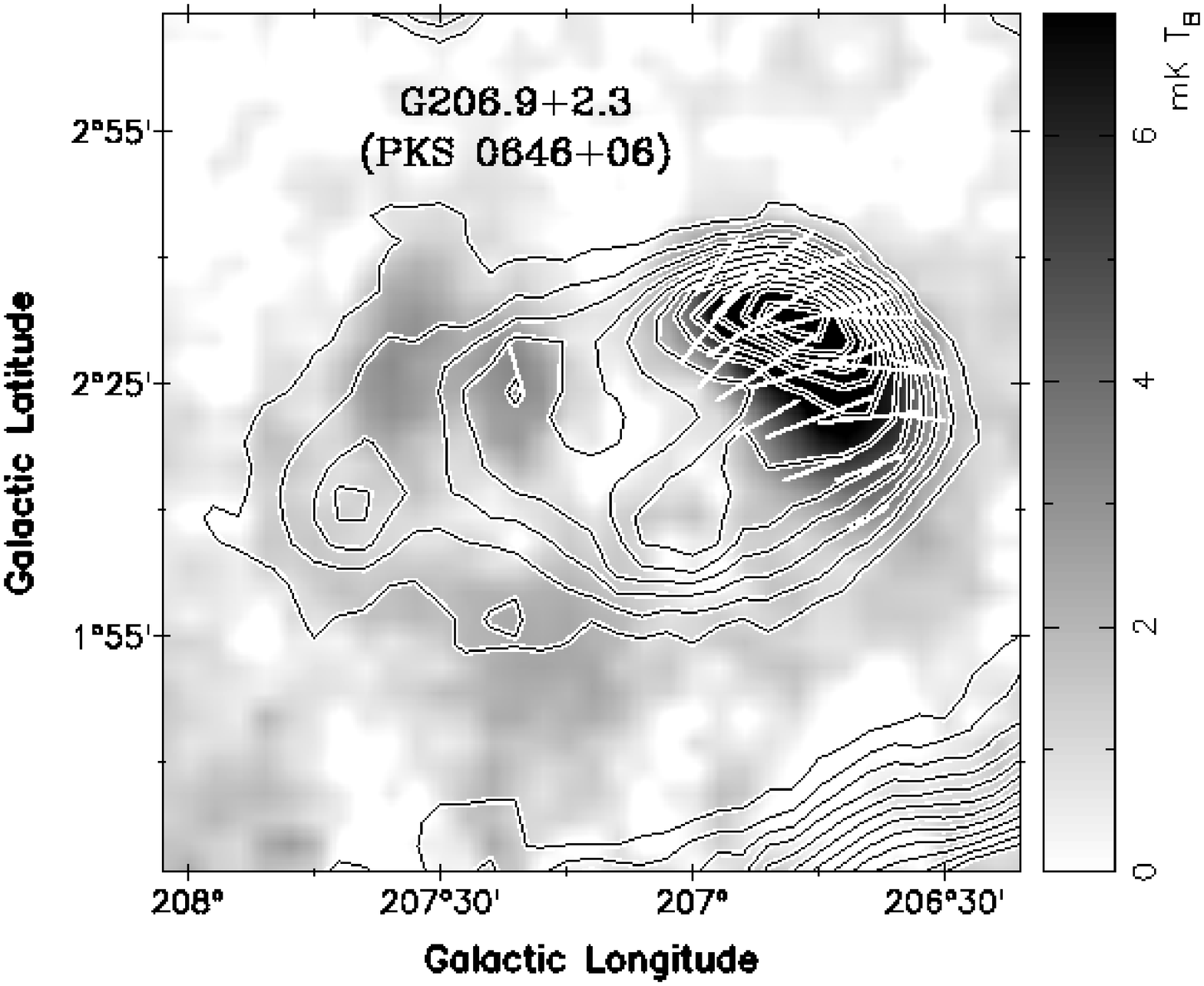}}\\
\end{center}
\caption{---continued.}
\end{figure*}
\addtocounter{figure}{-1}
\begin{figure*}
\begin{center}
\resizebox{0.49\textwidth}{!}{\includegraphics[angle=-90]{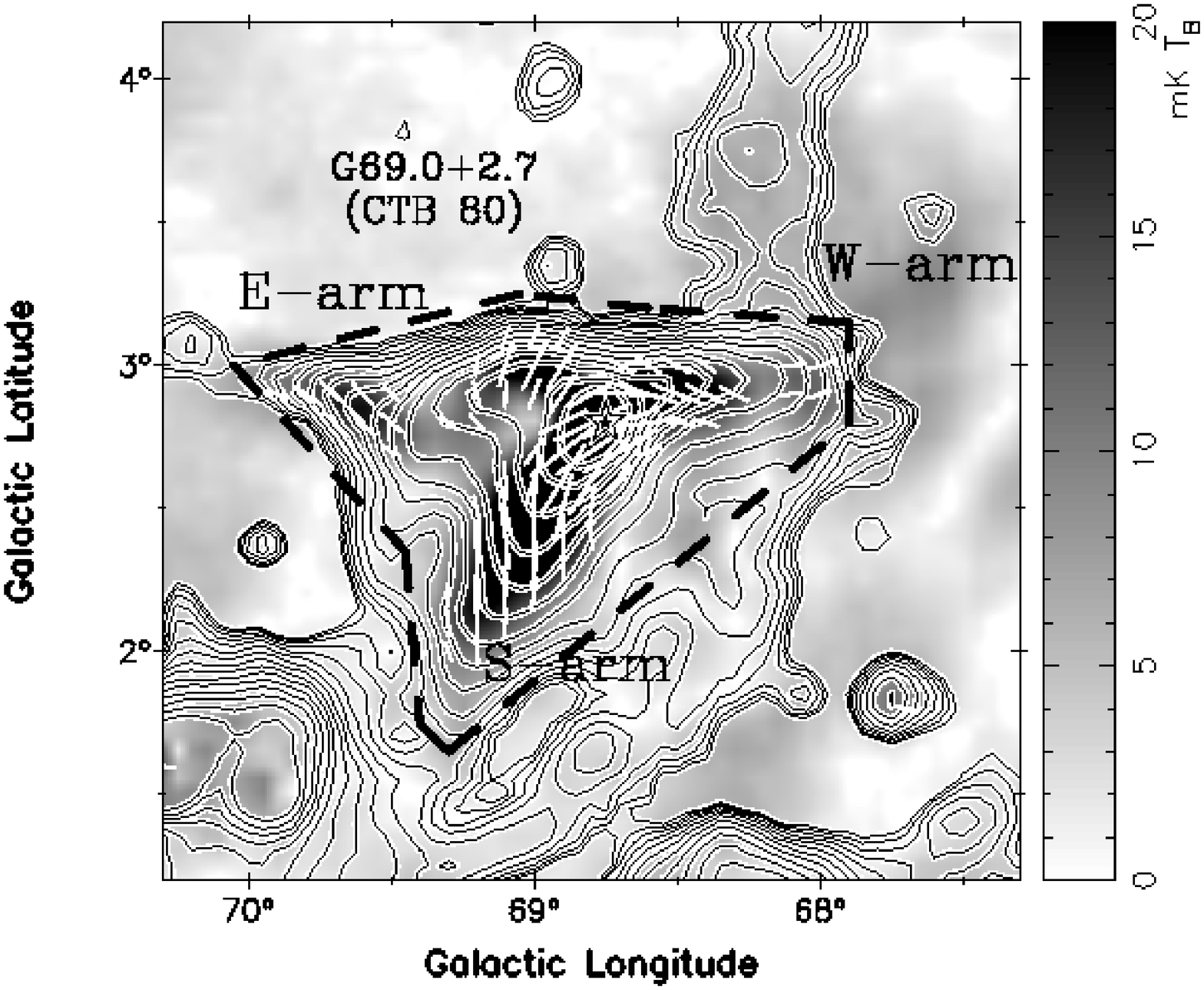}}
\resizebox{0.49\textwidth}{!}{\includegraphics[angle=-90]{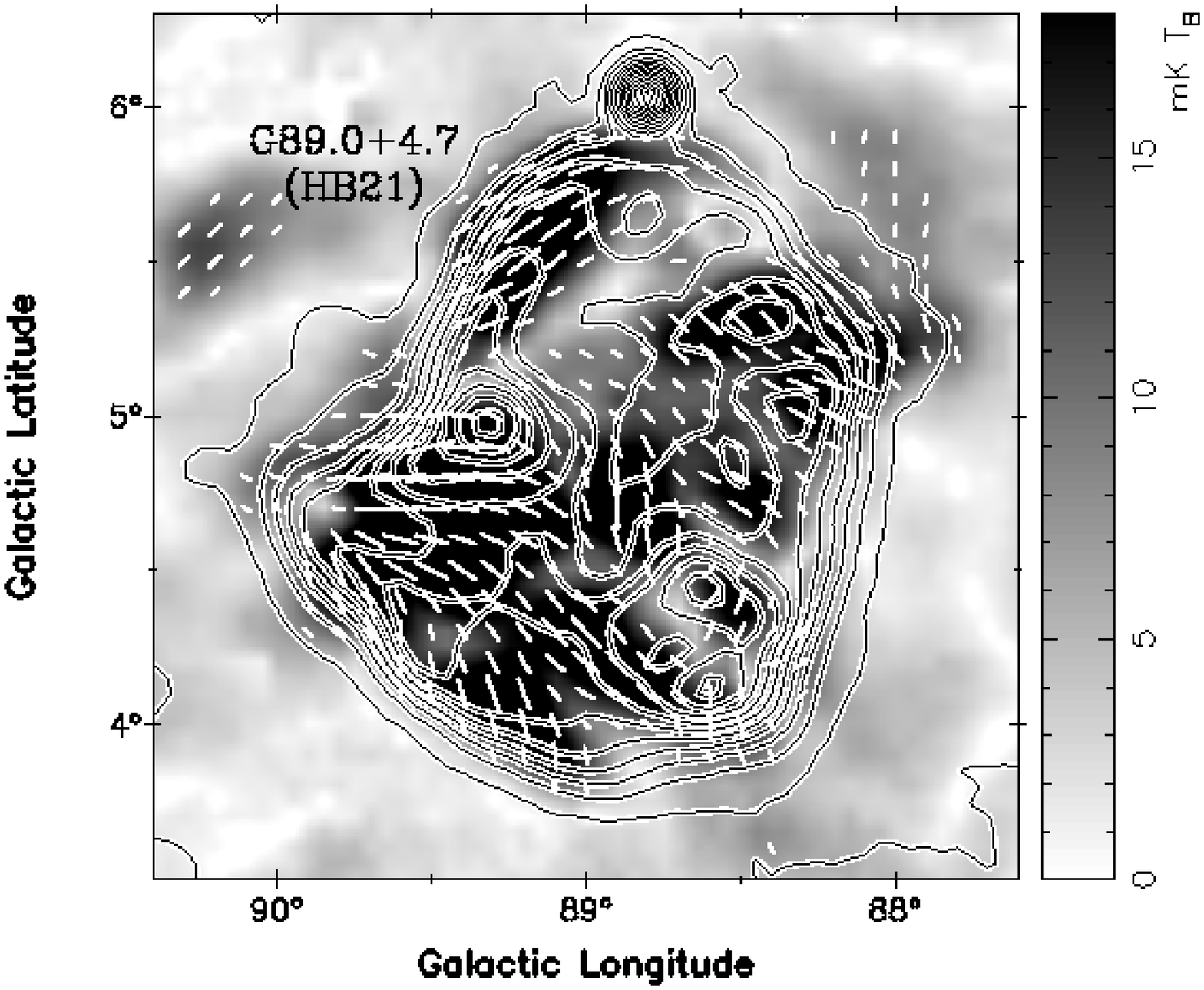}}\\
\resizebox{0.49\textwidth}{!}{\includegraphics[angle=-90]{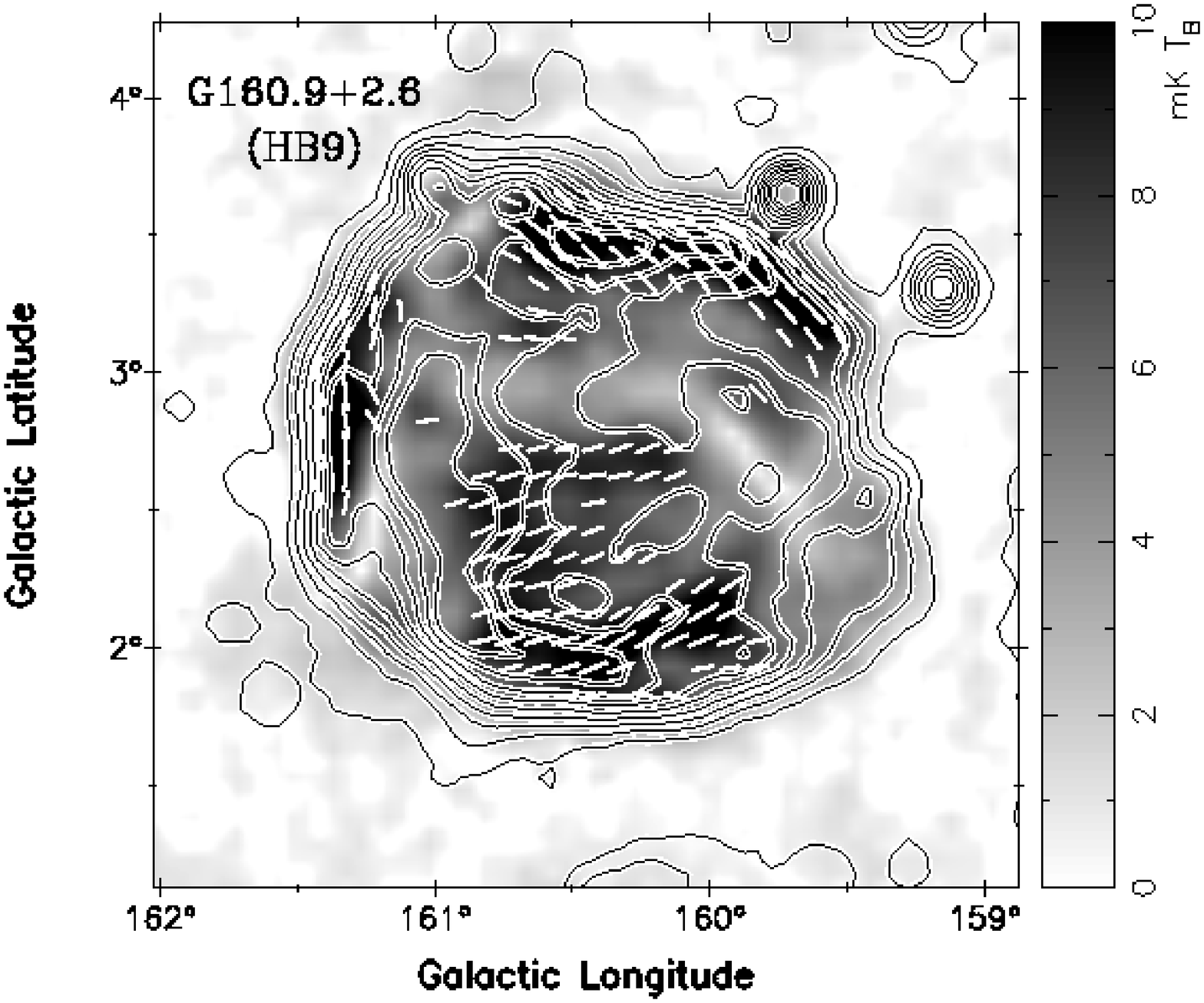}}
\resizebox{0.49\textwidth}{!}{\includegraphics[angle=-90]{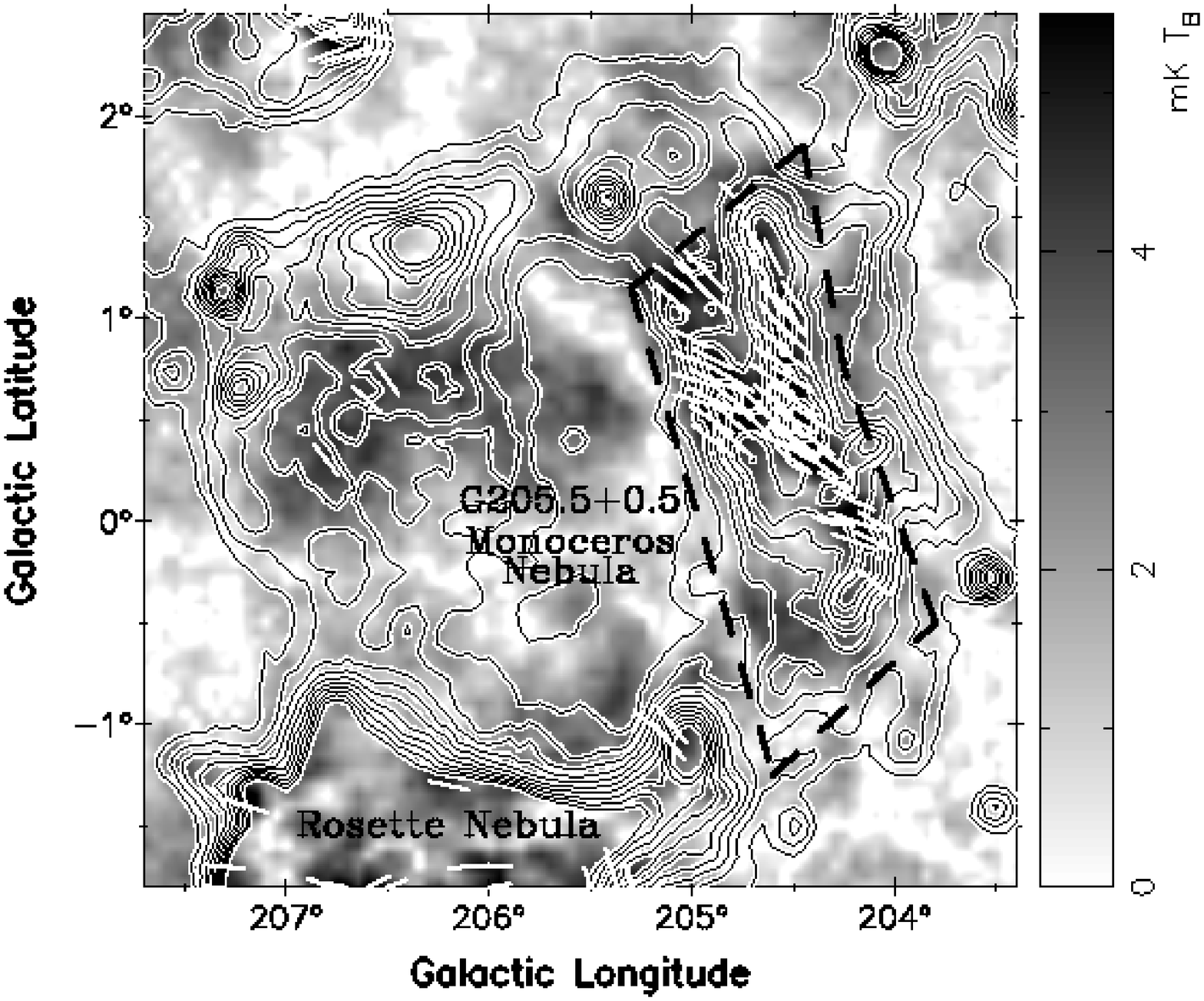}}
\end{center}
\caption{---continued.}
\label{SNRipi2}
\end{figure*}

\section{The $\lambda$6\ cm survey and the large SNRs}

The Sino-German $\lambda$6\ cm polarization survey of the Galactic
plane has been performed with a dual-channel $\lambda$6\ cm receiving
system constructed at MPIfR and mounted on the Urumqi 25-m radio
telescope. The receiving system had a system temperature of about $\rm
22~K\ T_{a}$, the central frequency was 4800~MHz, and the half power
beam width (HPBW) was $9\farcm5$. The observations were made by
scanning in both $\ell$ and $b$ directions. Stokes parameters $I$,
$U$, and $Q$ were stored simultaneously for each map. The system
set-up and the data reduction were described in detail by
\citet{Sun06, Sun07} and \citet{Gao10}.

The total intensity and polarization calibration was made with
observations of the primary polarization calibrator, 3C286 (flux
density at $\lambda$6\ cm $S_{\rm 6cm} = 7.5$~Jy, polarization angle
$PA = 33\degr$, and percentage polarization $\Pi = 11.3\%$), and the
secondary calibrators, 3C48 ($S_{\rm 6cm} = 5.5$~Jy, $PA = 108\degr$,
$\Pi = 4.2\%$) and 3C138 ($S_{\rm 6cm} = 3.9$~Jy, $PA = 169\degr$,
$\Pi = 10.8\%$). Instrumental effects in Stokes $U$ and $Q$ maps were
corrected for the leakage of total intensity $I$ into the polarization
channels as described by \citet{Sun07}.

The $\lambda$6\ cm survey region covers $10\degr \leq \ell \leq
230\degr$ and $|b| \leq 5\degr$ of the Galactic plane. According to
the SNR catalog of \citet{Green09}, twenty-two known SNRs in our
survey region have a size larger than $1\degr$. In the $\lambda$6\ cm
survey, the size of G43.9+1.6 is less than $1\degr$, while G127.1+0.5,
G166.0+4.3 (VRO 42.05.01), and G189.1+3.0 (IC443) are larger than
$1\degr$. Thus 24 SNRs are left.  Four SNRs, SNR G126.2+1.6 and
G127.1+0.5 \citep{Sun07}, SNR G180.0$-$1.7 \citep{Xiao08}, and
G130.7+3.1 \citep{Shi08a}, have previously been studied. SNRs
G13.3$-$1.3 is omitted in this paper because it was not reliably
detected in the $\lambda$6\ cm survey. G28.8+1.5 is omitted because
only a small section of the radio shell is detected and is confused by
strong diffuse Galactic emission.  Faint diffuse emission is detected
in the area of SNR G32.0$-$4.9 and only some bright knots above
3$\sigma$ noise level were seen in the survey map. We cannot
unambiguously associate the diffuse emission with the SNR and
therefore do not study this object.  SNR G108.2$-$0.6 \citep{Tian07}
was also skipped because it is confused by strong thermal emission
from nearby \ion{H}{II} regions and cannot be separated. However, the
two SNRs, G82.2+5.3 (W63) and G89.0+4.7 (HB21), are located at the $b
= +5\degr$ boundary of the $\lambda$6\ cm survey and have been
completely observed \citep{Xiao11}. They were included. We therefore
study 17 objects in this paper, where G192.8$-$1.1 is disproved as
being a SNR in Sect~3.1.

%
\begin{figure*}[tbh]
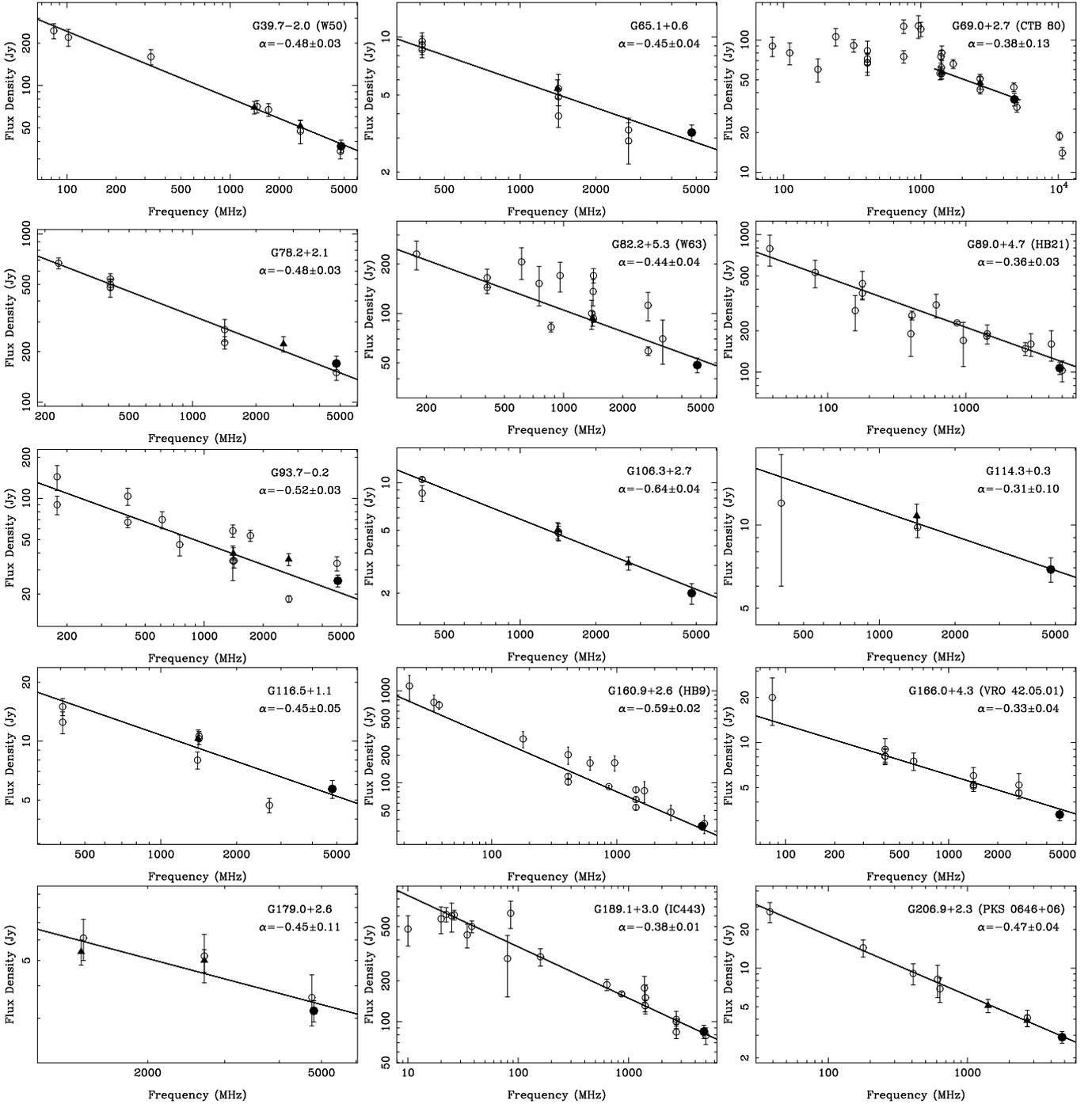

\begin{center}
\resizebox{0.32\textwidth}{!}{\includegraphics[angle=-90]{16311fg2a.ps}}
\resizebox{0.32\textwidth}{!}{\includegraphics[angle=-90]{16311fg2b.ps}}
\resizebox{0.32\textwidth}{!}{\includegraphics[angle=-90]{16311fg2c.ps}}\\
\resizebox{0.32\textwidth}{!}{\includegraphics[angle=-90]{16311fg2d.ps}}
\resizebox{0.32\textwidth}{!}{\includegraphics[angle=-90]{16311fg2e.ps}}
\resizebox{0.32\textwidth}{!}{\includegraphics[angle=-90]{16311fg2f.ps}}\\
\resizebox{0.32\textwidth}{!}{\includegraphics[angle=-90]{16311fg2g.ps}}
\resizebox{0.32\textwidth}{!}{\includegraphics[angle=-90]{16311fg2h.ps}}
\resizebox{0.32\textwidth}{!}{\includegraphics[angle=-90]{16311fg2i.ps}}\\
\resizebox{0.32\textwidth}{!}{\includegraphics[angle=-90]{16311fg2j.ps}}
\resizebox{0.32\textwidth}{!}{\includegraphics[angle=-90]{16311fg2k.ps}}
\resizebox{0.32\textwidth}{!}{\includegraphics[angle=-90]{16311fg2l.ps}}\\
\resizebox{0.32\textwidth}{!}{\includegraphics[angle=-90]{16311fg2m.ps}}
\resizebox{0.32\textwidth}{!}{\includegraphics[angle=-90]{16311fg2n.ps}}
\resizebox{0.32\textwidth}{!}{\includegraphics[angle=-90]{16311fg2o.ps}}\\
\caption{Flux densities at various frequencies were used to derive the
  integrated radio spectra for 15 SNRs (not including G205.5+0.5). The
  black dots indicate the newly derived flux density at $\lambda$6\ cm
  from the Urumqi observations, and the triangles for the newly
  derived flux densities from the Effelsberg $\lambda$11\ cm and
  $\lambda$21\ cm surveys. The open circles are integrated flux
  densities taken from literature (see references in Table~1.)}
\label{spectrum_SNR}
\vspace{-2mm}
\end{center}
\end{figure*}
%
\begin{sidewaystable*}[p]
\vspace{180mm}
\caption{Integrated flux densities and polarization intensities for 15
  large SNRs (not including G205.5+0.5, see text) derived from the
  Urumqi $\lambda$6\ cm survey. The integrated flux densities at
  $\lambda$11\ cm and $\lambda$21\ cm newly extracted from the
  Effelsberg surveys are also listed. The newly derived spectral
  indices and the previously published flux densities at
  $\lambda$6\ cm, spectral indices of SNRs and their references are
  listed for comparison. References for flux densities at other
  frequencies are given in the last column.}
\label{sources}
\vspace{-1mm}
\centering
\renewcommand{\footnoterule}{}
\begin{tabular}{lrrrrcrcccl}
\hline\hline
\multicolumn{1}{c}{SNR name} &\multicolumn{1}{c}{$\rm S_{\lambda6\ cm}$}  & \multicolumn{1}{c}{$\rm PI_{\lambda6\ cm}$} &\multicolumn{1}{c}{$\rm S_{\lambda11\ cm}$} &\multicolumn{1}{c}{$\rm S_{\lambda21\ cm}$} &New $\alpha$ &Prev. $\rm S_{\lambda6\ cm}$ &Ref. &Prev. $\alpha$ &Ref. & Ref. for $S_{\nu}$ \\
         & \multicolumn{1}{c}{(Jy)}                 & \multicolumn{1}{c}{(Jy)}         & \multicolumn{1}{c}{(Jy)}      & \multicolumn{1}{c}{(Jy)}      &             & (Jy)                      &     &               &   & at other freq. ($\nu$) \\
\hline
G39.7$-$2.0 (W50)      &$37.0\pm3.8$  &$5.3\pm0.6$  &$51.5\pm5.2$&$69.7\pm7.1$ &$-0.48\pm0.03$  &$34\pm4$      &2         &$-0.48\pm0.03$  &3  &1,2,3,4 \\
G65.1+0.6              & $3.2\pm0.3$  &$0.8\pm0.1$  &$\cdots$    &$5.4\pm0.6$  &$-0.45\pm0.04$  &$\cdots$      &$\cdots$  &$-0.61\pm0.09$  &6  &5,6,7 \\
G69.0+2.7 (CTB~80)     &$35.6\pm3.9$  &$4.7\pm0.5$  &$47.6\pm5.0$&$56.7\pm6.3$ &$\cdots$        &$44.0\pm3.3$  &9         &$-0.45\pm0.03$  &6  &6,8,9,10 \\
G78.2+2.1              &$170\pm18$    &$1.2\pm0.1$  &$222\pm23$  &$\cdots$     &$-0.48\pm0.03$  &$150\pm15$    &11        &$-0.51\pm0.03$  &6  &6,11,12,13,14 \\   
G82.2+5.3 (W63)        &$48.5\pm4.9$  &$7.5\pm0.8$  &$\cdots$    &$93.1\pm9.5$ &$-0.44\pm0.04$  &$>38.5\pm4.0$ &15        &$-0.48\pm0.04$  &6  &6,15,16 \\
G89.0+4.7 (HB21)       &$107\pm11$    &$11.3\pm1.1$ &$\cdots$    &$\cdots$     &$-0.36\pm0.03$  &$103\pm18$    &17        &$-0.38\pm0.03$  &6  &6,16,17,18 \\
G93.7$-$0.2            &$25.0\pm2.5$  &$5.2\pm0.5$  &$35.9\pm3.6$&$39.6\pm4.0$ &$-0.52\pm0.03$  &$33.5\pm4.0$  &14        &$-0.65\pm0.03$  &6  &6,19,20,21,22 \\
G106.3+2.7             &$2.0\pm0.3$   &$0.4\pm0.1$  &$3.1\pm0.3$ &$5.0\pm0.6$  &$-0.64\pm0.04$  &$\cdots$      &$\cdots$  &$-0.61\pm0.07$  &6  &6,23 \\ 
G114.3+0.3             &$6.9\pm0.7$   &$0.9\pm0.1$  &$\cdots$    &$10.8\pm1.1$ &$-0.31\pm0.10$  &$\cdots$      &$\cdots$  &$-0.49\pm0.25$  &6  &24 \\
G116.5+1.1             &$5.7\pm0.6$   &$1.6\pm0.2$  &$\cdots$    &$10.3\pm1.1$  &$-0.45\pm0.05$  &$\cdots$      &$\cdots$  &$-0.53\pm0.08$  &6  &6,24,25 \\
G160.9+2.6 (HB9)       &$33.9\pm3.4$  &$5.3\pm0.5$  &$\cdots$    &$\cdots$     &$-0.59\pm0.02$  &$36\pm8$      &26        &$-0.64\pm0.02$  &6  &6,16,26,27,28 \\
G166.0+4.3 (VRO~42.05.01) &$3.3\pm0.3$&$0.6\pm0.1$  &$\cdots$    &$\cdots$     &$-0.33\pm0.04$  &$\cdots$      &$\cdots$  &$-0.37\pm0.03$  &6  &6,18,29 \\
G179.0+2.6             &$3.2\pm0.3$   &$0.4\pm0.1$  &$5.0\pm0.5$ &$5.4\pm0.6$  &$-0.45\pm0.11$  &$3.6\pm0.8$   &30        &$-0.30\pm0.15$  &30 &30 \\
G189.1+3.0 (IC443)     &$84.6\pm9.4$  &$2.6\pm0.3$  &$\cdots$    &$\cdots$     &$-0.38\pm0.01$  &$79\pm11$     &31        &$-0.36\pm0.02$  &31 &16,31 \\ 
G206.9+2.3 (PKS~0646+06)&$2.9\pm0.3$  &$0.4\pm0.1$  &$3.9\pm0.4$ &$5.1\pm0.6$  &$-0.47\pm0.04$  &$\cdots$      &$\cdots$  &$-0.45\pm0.03$  &32 &32\\

\hline
\end{tabular}
\vspace{-1mm}
\tablebib{
(1) \citet{Downes81}, (2) \citet{Downes86}, (3) \citet{Dubner98}, (4) \citet{Kovalenko94}, 
(5) \citet{Landecker90}, (6) \citet{Kothes06b}, (7) \citet{Tian06x}, (8) \citet{Castelletti03}, 
(9) \citet{Mantovani85}, (10) \citet{Velusamy76}, (11) \citet{Wendker91}, (12) \citet{Higgs77b}, (13) \citet{Pineault90}, 
(14) \citet{Zhang97}, (15) \citet{Higgs91}, (16) \citet{Reich03}, (17) \citet{Hirabayashi72}, 
(18) \citet{Willis73}, (19) \citet{Velusamy74}, (20) \citet{Mantovani82}, (21) \citet{Landecker85}, 
(22) \citet{Mantovani91}, (23) \citet{Pineault00}, (24) \citet{Tian06y}, (25) \citet{Reich81}, 
(26) \citet{Dwarakanath82}, (27) \citet{Roger99}, (28) \citet{Leahy07}, (29) \citet{Leahy05}, 
(30) \citet{Fuerst86}, (31) \citet{Erickson85}, (32) \citet{Graham82}.}
{\vspace{-1mm}
}
\end{sidewaystable*}

\section{Results and discussions}

The SNR images were obtained after we filtered out the large-scale
Galactic diffuse emission by using the technique of ``background
filtering'' \citep{Sofue79}. A twisted hyper-plane was defined by the
average values of pixels without any obvious structure in the four map
corners and subtracted for each SNR to remove the local-background
Galactic diffuse emission in $I$, $U$, and $Q$ maps. We obtained total
intensity and polarization images of these SNRs as shown in
Fig.~\ref{SNRipi}.

We integrated the flux density ($S_{\rm 6cm}$) and polarization
intensity ($PI_{\rm 6cm}$) of SNRs within the boundary defined by the
3$\sigma$ noise level around each SNR in total intensity (see Table
1), where $\sigma$ is the rms of the map estimated from the
surrounding areas of each SNR without obvious structure. For the SNRs
G39.7$-$2.0 (W50), G65.1+0.6, G78.2+2.1, G82.2+5.3 (W63), G106.3+2.7,
G189.1+3.0 (IC443), and G69.0+2.7 (CTB~80), diffuse emission adjacent
to the SNR is detected in the $\lambda$6\ cm map. We thus outline the
SNR boundary according to published radio maps at other frequencies
for flux density integration (see Fig.~\ref{SNRipi}) to avoid possible
confusion. For the complicated structure of SNR G205.5+0.5 (Monoceros
Nebula), we outline only the western shell for the integration at
$\lambda$6\ cm, $\lambda$11\ cm, and $\lambda$21\ cm to determine its
spectrum. The uncertainty of each integrated SNR flux density is
calculated from the rms of a map and the pixel number of the
integration area. The calibration and base-level uncertainties are
assumed to be 10\%, and added to the final value of the
uncertainty. Compact sources exceeding a flux density of 20~mJy in the
NVSS catalogue at 1.4~GHz \citep{Condon98} and located within the SNR
boundary were discounted from the integrated SNR flux densities. The
flux densities of point-like sources at $\lambda$6\ cm were estimated
from the extrapolation of the NVSS flux density at 1.4~GHz
\citep{Condon98} using a spectral index either taken from
\citet{Vollmer05} or derived from the flux densities at 408~MHz and
1420~MHz from the Canadian Galactic Plane Survey (CGPS)
\citep{Taylor03}.  If no spectral index is available for a source, we
adopted the mean value of $\alpha = -0.9$ according to \citet{Zhang03}
obtained from the NVSS and the WSRT source samples.

We obtained integrated flux densities of a number of SNRs (see Table
1) from the Effelsberg $\lambda$11\ cm survey \citep{Fuerst90,
  Reich9011}, $\lambda$21\ cm survey \citep{Reich9021, Reich97}, and
the $\lambda$21\ cm Effelsberg Medium Latitude Survey (EMLS)
\citep{Uyaniker99} within the same boundary as used for the
$\lambda$6\ cm integration.

Using the newly determined flux densities at $\lambda$6\ cm,
$\lambda$11\ cm, and $\lambda$21\ cm, together with previously
published integrated flux densities at other frequencies (see
Table~1), we determined the spectral indices $\alpha$ (here $S_{\nu}
\sim \nu^{\alpha}$) of 15 SNRs, except for SNR G205.5+0.5, where $\nu$
is the observing frequency. Flux densities and the determined spectra
are shown in Fig.~\ref{spectrum_SNR}. The spectra of a few SNRs were
verified by using the TT-plot method \citep{Turtle62}, i.e. the
brightness temperatures $T_B$ at two frequencies are plotted against
each other. The brightness temperature spectral index, $\beta$, is
defined as $T_B = \nu^{\beta}$. In general, $\beta = \alpha - 2$.

\subsection{G192.8$-$1.1 is not a SNR}

G192.8$-$1.1 (PKS 0607+17) has been proposed as a SNR with an angular
extent of about $80\arcmin$ \citep{Milne69, Caswell70}. The
non-thermal spectral index derived was $\alpha \sim -0.5$
\citep{Milne74, Dickel75}. \citet{Berkhuijsen74} suggested that
G192.8$-$1.1 is a small part of an even larger SNR, which she called
the `Origem Loop' with a diameter of about 5$\degr$, but later
\citet{Caswell85} performed high angular resolution ($\sim 1\arcmin$)
observations at $\lambda$21\ cm using the DRAO synthesis telescopes
and argued again that G192.8$-$1.1 is a discrete SNR with a radio
spectral index of $\alpha \sim -0.55$. However, previous measurements
were limited either by low sensitivity or too coarse angular
resolution, which could lead to an incorrect conclusion.

\begin{figure}
\begin{center}
\includegraphics[angle=-90, width=0.45\textwidth]{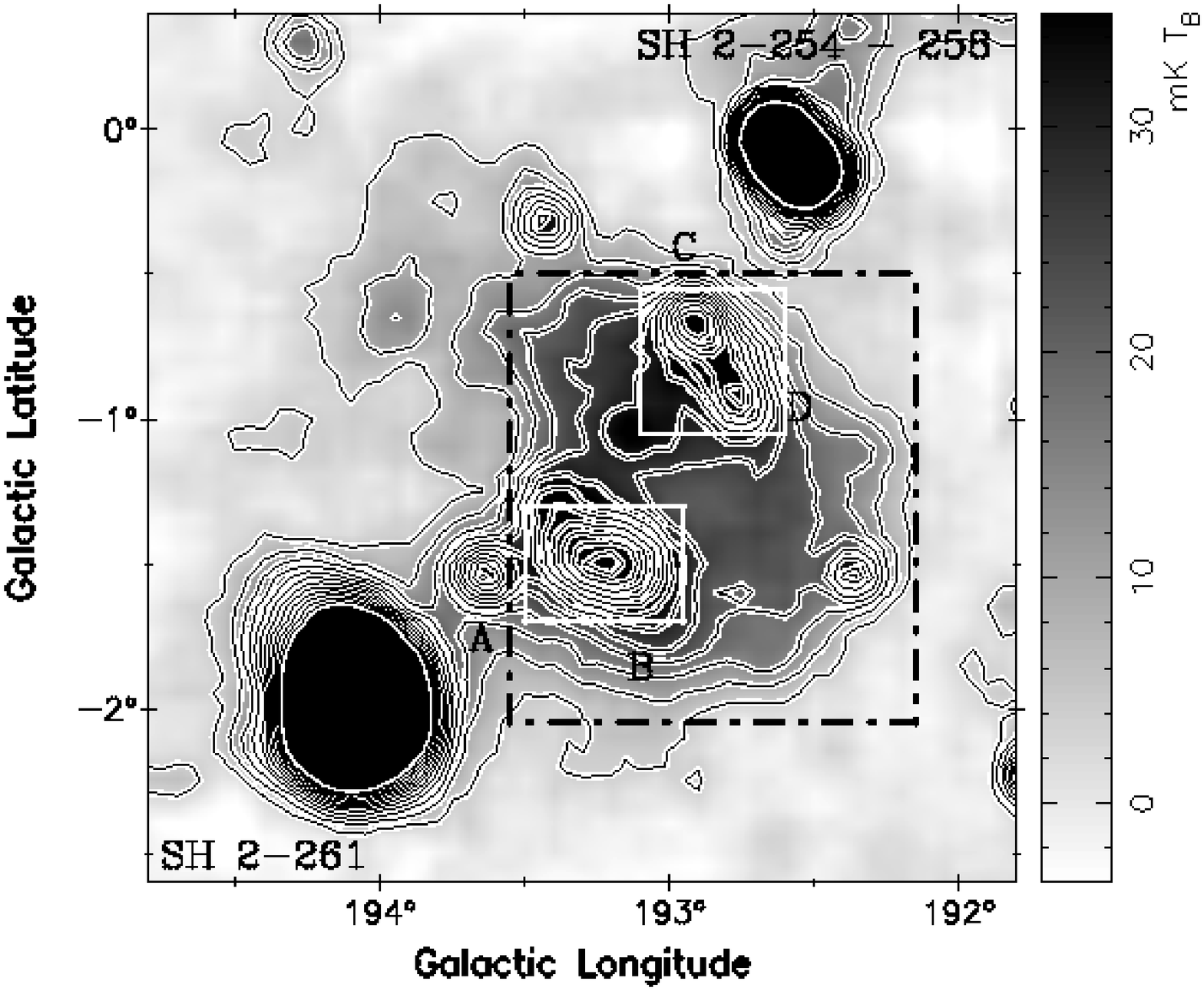}
\includegraphics[angle=-90, width=0.45\textwidth]{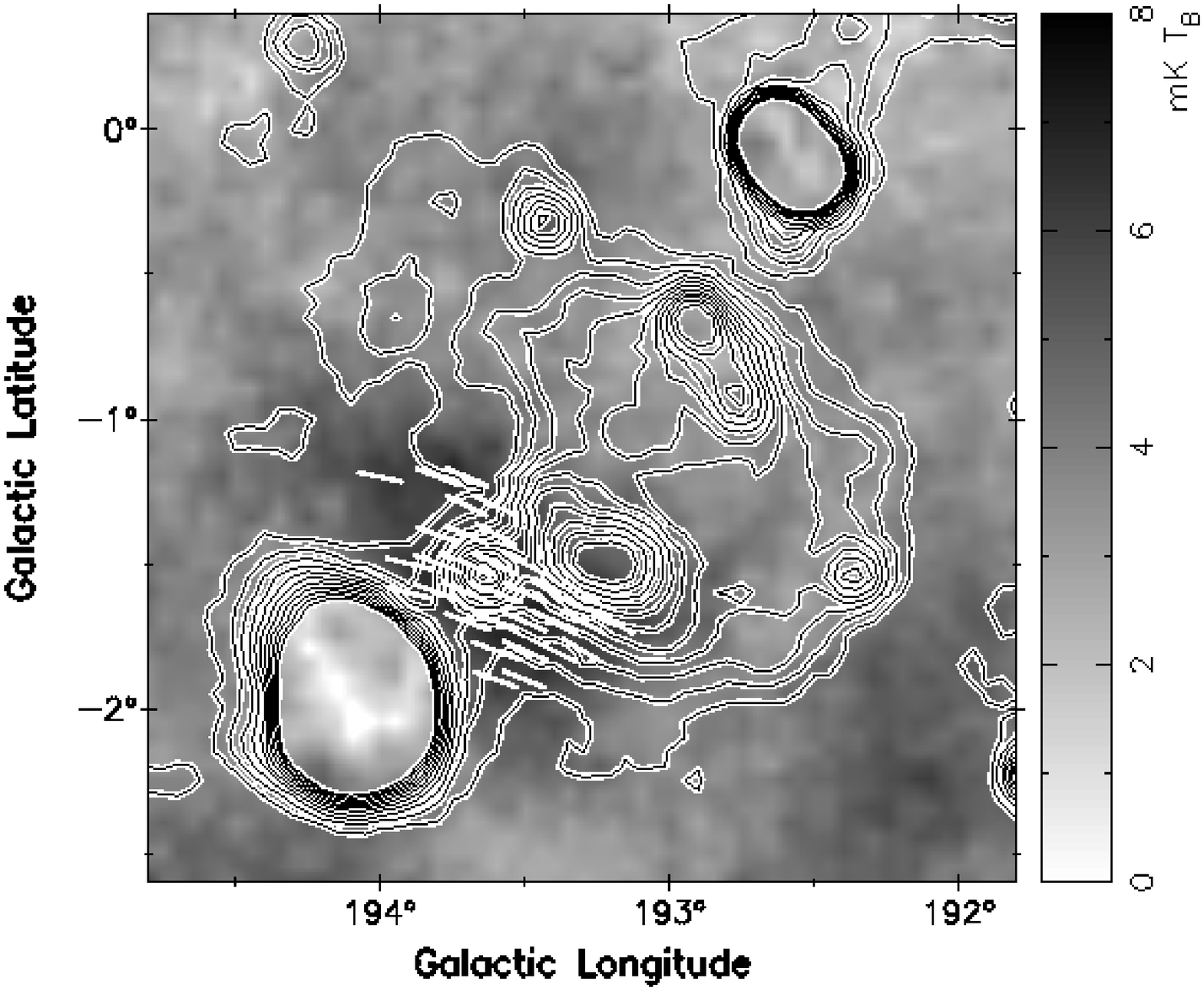}
\includegraphics[angle=-90, width=0.45\textwidth]{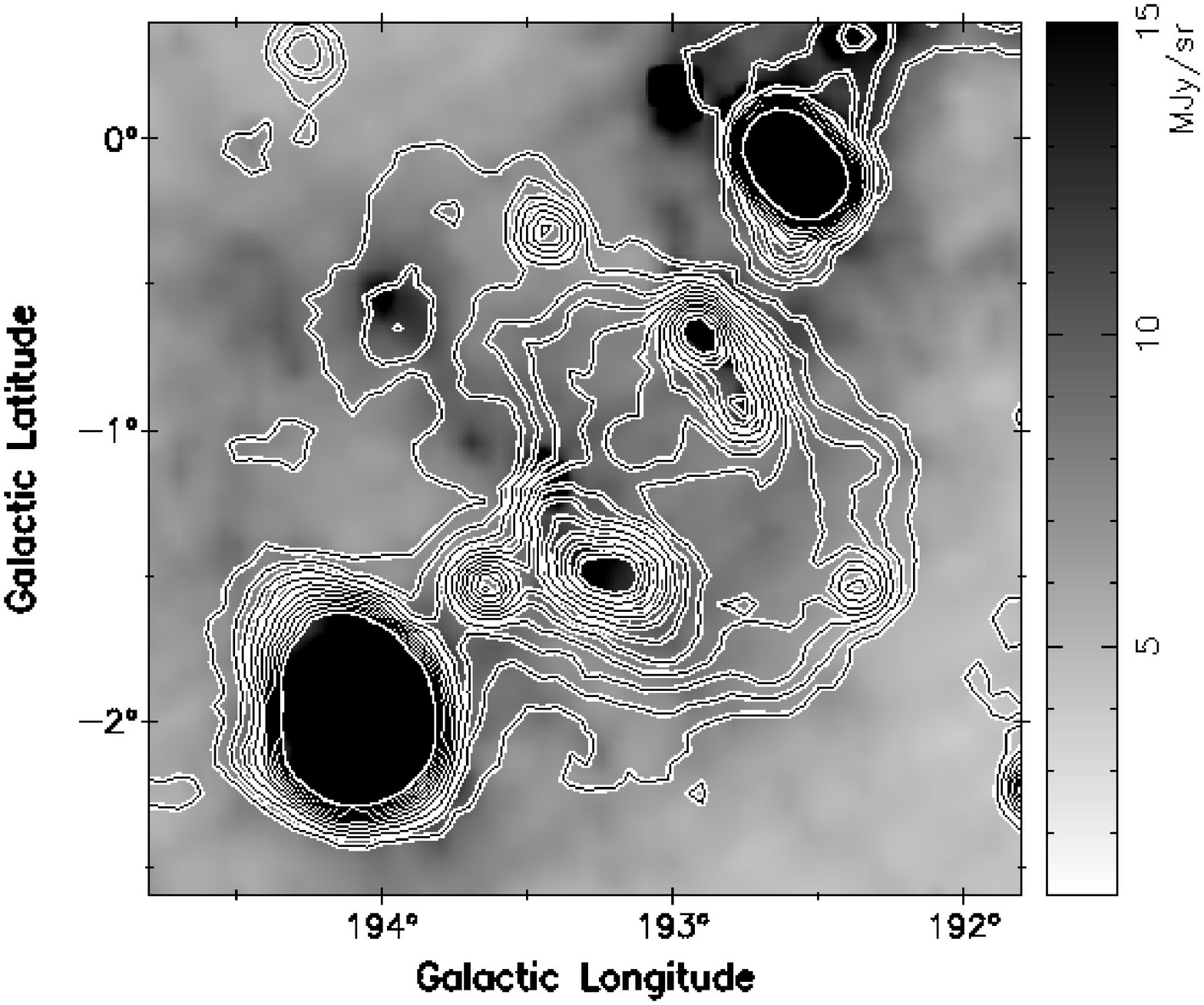}
\caption{{\it Top}: Total intensity ($I$) image at $\lambda$6\ cm
  for G192.8$-$1.1 (PKS 0607+17) and its surroundings. The
  ``background filtering'' technique was used to remove large-scale
  emission. Contours start at $\rm 3.0~mK\ T_{B}$ ($3\sigma$ level)
  and run in steps of $\rm 5.0~mK\ T_{B}$. The big rectangle area
  includes the G192.8$-$1.1, which was used for the TT-plot analysis
  by excluding the two areas marked by white small rectangles. {\it
    Middle}: PI image of G192.8$-$1.1 area (PKS 0607+17) at
  $\lambda$6\ cm.  The same total intensity contours are overlaid as
  in the {\it Top} plot. Bars are plotted for observed B-vectors
  (i.e. $PA+90\degr$) at $\lambda$6\ cm if $PI \geqslant \rm
  6~mK\ T_{B}$.  {\it Bottom}: Infrared image at $60\mu$m of the same
  area \citep{Cao97} overlaid by $\lambda$6\ cm $I$ contours.}
\label{G192}
\end{center}
\end{figure}

\begin{figure}[bth]
\begin{center}
\includegraphics[angle=-90, width=0.4\textwidth]{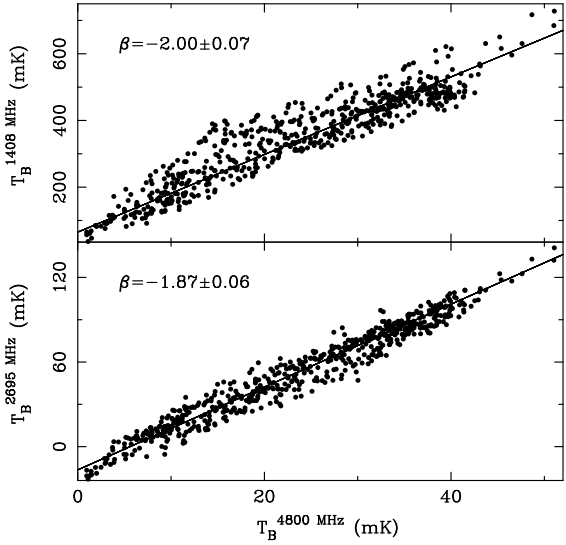}
\caption{TT-plot for the G192.8$-$1.1 ``plateau'' using the Urumqi
  $\lambda$6\ cm and the Effelsberg $\lambda$21\ cm (1408~MHz) survey
  data ({\it top panel}) and the $\lambda$6\ cm and the Effelsberg
  $\lambda$11\ cm (2695~MHz) survey data ({\it bottom panel}).}
\label{G192_tt}
\end{center}
\end{figure}

Sensitive survey observations from Urumqi at $\lambda$6\ cm
\citep{Gao10}, Effelsberg at $\lambda$11\ cm \citep{Fuerst90}, and
$\lambda$21\ cm \citep{Reich97} all covered this object. The
$\lambda$6\ cm total intensity map shows G192.8$-$1.1 in the vicinity
of a number of bright \ion{H}{II} regions (Fig.~\ref{G192}). The
\ion{H}{II} region to the southeast, SH 2-261, has a distance of 1~kpc
\citep{Chava87}; the cluster of \ion{H}{II} regions, SH 2-254 to 258,
to the northwest have distances of about 2.5~kpc
\citep{Carpenter95}. Four individual objects, A ($\ell = 193\fdg60,
b=-1\fdg50$), B ($\ell = 193\fdg20, b = -1\fdg50$), C ($\ell =
192\fdg90, b = -0\fdg60$), and D ($\ell = 192\fdg80, b = -0\fdg90$) as
indicated in Fig.~\ref{G192} were discussed by \citet{Day72} based on
their 2.7~GHz Galactic survey data obtained with the Parkes telescope.
All of them except source A were believed to be parts of the SNR. The
point-like source A was listed as an \ion{H}{II} region by
\citet{Paladini03} according to \citet{Felli72}. However, this source
also named as 4C 16.15 clearly has a non-thermal spectrum
\citep{Vollmer05} and is therefore not an \ion{H}{II} region. The
elongated source B is located west of source A. \citet{Caswell85}
considered source B as part of the SNR shell of G192.8$-$1.1.  We
found that source B has a thermal spectrum according to the TT-plot
between the Urumqi $\lambda$6\ cm and the Effelsberg $\lambda$21\ cm
survey data.  The brightness-temperature spectral index is $\beta =
-2.06\pm0.06$.  In addition, strong infrared
emission (see Fig.~\ref{G192}, bottom plot) is visible, which proves
its thermal nature.

The enhanced radio emission region where the source C and D reside is
regarded as another part of the SNR shell of G192.8$-$1.1
\citep{Caswell85}.  Source C consists of a few point-like sources
including the \ion{H}{II} region SH 2-259. It appears slightly
extended towards source D at $9\farcm5$ angular resolution. Source D
was resolved into two sources in the NVSS, and its flux density mainly
originates in the source NVSS J060951+172532, which has a spectral
index of $\alpha \sim -0.73$ \citep{Vollmer05} and is most likely
extragalactic. In summary, sources A, B, C, and D do not seem to be
related to the proposed SNR.

The diffuse emission plateau was regarded by \citet{Caswell85} as part
of the SNR. We subtracted all point-like sources including source A
and discarded the regions where sources B, C, and D are located. Using
the TT-plot method, we found a well-defined flat spectrum with a slope
of $\beta = -2.00\pm0.07$ between the Urumqi $\lambda$6\ cm and the
Effelsberg $\lambda$21\ cm survey data, and $\beta = -1.87\pm0.06$
between $\lambda$6\ cm and Effelsberg $\lambda$11\ cm survey data (see
Fig.~\ref{G192_tt}). Taking the diffuse infrared emission in this area
\citep{Cao97} into account, we conclude that the extended emission 
plateau is of thermal nature.

The diffuse polarization patch in Fig.~\ref{G192} located between
G192.8$-$1.1 and the \ion{H}{II} region SH 2-261 is unrelated to
G192.8$-$1.1.

To summarize, we found a thermal spectrum for the diffuse plateau of
G192.8$-$1.1, which is in contradiction to the non-thermal spectrum
reported in previous investigations. The extended source in the
southeast, which was regarded as part of the SNR shell, is also
identified as a thermal source. No polarized emission from
G192.8$-$1.1 was detected at $\lambda$6\ cm. This ``object'',
G192.8$-$1.1, is evidently not a SNR, because it consists of several
unresolved \ion{H}{II} regions, background radio sources, and a
thermal emission plateau.

\subsection{G39.7$-$2.0 (W50)}

W50 is an extended SNR located at about 5.5~kpc \citep{Blundell04,
  Lockman07}.  It consists of a central component with a diameter of
about 1$\degr$ and two ``ears'' extended to the northwest and
southeast (see Fig.~\ref{SNRipi}).  The spectacular X-ray binary
system SS~433 was located at the center of the SNR
\citep[e.g.][]{Brinkmann96}. Two relativistic jets from SS~433 are
precessing around the major axis of W50. According to
\citet{Begelman80}, the ``ears'' of W50 are caused by the pressure of
these two jets and interact with the ambient \ion{H}{I}
\citep{Lockman07}. \citet{Jowett95} mapped SS~443 at 5~GHz with the
MERLIN array and detected ejected radio blobs moving away from
SS~443. \citet{Downes81} and \citet{Downes86} observed the total
intensity and the polarized emission of W50 at $\lambda$18\ cm,
$\lambda$11\ cm, and $\lambda$6\ cm with the 100-m Effelsberg radio
telescope. The spectral index and $RM$ distributions of this SNR were
studied in detail.

The Urumqi $\lambda$6\ cm survey clearly shows the SNR W50 with strong
polarized emission along its northeastern shell of the central
part. SS~433 is seen at the center, but the related radio jets
\citep{Jowett95} cannot be resolved. We obtained integrated flux
densities of $S_{\rm 6cm} = 37.0\pm3.8$~Jy from the Urumqi survey
data, $S_{\rm 11cm} = 51.5\pm$5.2~Jy and $S_{\rm 21cm} =
69.7\pm7.1$~Jy from the Effelsberg $\lambda$11\ cm and $\lambda$21\ cm
survey data. All of these agree with the values obtained by
\citet{Downes81, Downes86} and \citet{Dubner98}. Using these flux
densities together with previously published integrated flux
densities, we derived an overall spectral index of $\alpha =
-0.48\pm0.03$, consistent with that found by \citet{Dubner98}. The
TT-plot using the Urumqi $\lambda$6\ cm and Effelsberg $\lambda$11\ cm
survey data gave $\beta = -2.41\pm0.01$ for the central part, $\beta =
-2.52\pm0.02$ for the southeast ``ear'', and $\beta = -2.57\pm0.14$ for
the northwest ``ear'', respectively. The central part appears to have
a slightly flatter spectrum than the southeast ``ear'', confirming the
finding of \citet{Downes86}.

\subsection{G65.1+0.6}

G65.1+0.6 was first identified as a faint shell-type SNR by
\citet{Landecker90} from 408~MHz and 1420~MHz observations with the
DRAO Synthesis Telescope. It shows strong southern shell emission,
while its emission in the north is rather diffuse. \citet{Tian06x}
analyzed CGPS \ion{H}{I} data and proposed a SNR distance of
9.2~kpc. Polarization from the southern shell was already observed at
$\lambda$6\ cm with the Effelsberg 100-m telescope
\citep{Seiradakis85}. Despite a smaller beam, no polarized emission
was detected with the Effelsberg 100-m telescope at $\lambda$11\ cm
\citep{Duncan99} indicating significant depolarization at that
wavelength. This is not unexpected in view of the large distance of
the SNR.

Our $\lambda$6\ cm data confirm the oval shape of G65.1+0.6 (see
Fig.~\ref{SNRipi}). We obtained an integrated flux density at
$\lambda$6\ cm of $S_{\rm 6cm} = 3.2\pm0.3$~Jy. We also calculated the
$\lambda$21\ cm integrated flux density from the Effelsberg
$\lambda$21\ cm survey to be $S_{\rm 21cm} = 5.4\pm0.6$~Jy. This value
is consistent with those quoted by \citet{Landecker90} and
\citet{Tian06x} from the CGPS, which used the Effelsberg
$\lambda$21\ cm survey for missing zero-spacings. using these flux 
densities together with the
flux densities at 408~MHz \citep{Landecker90, Kothes06b, Tian06x} and
other bands (see references in Table~1), we obtained a radio spectral
index of $\alpha = -0.45\pm0.04$ (see Fig.~\ref{spectrum_SNR}).

We detected very weak polarization at $\lambda$6\ cm (see
Fig.~\ref{SNRipi}), which seems to be related to the SNR shell, as the
magnetic field direction is almost tangential.

\subsection{G69.0+2.7 (CTB~80)}

CTB~80 is a SNR with an enigmatic shape. It consists of a bright
pulsar wind nebula (PWN) in its center, a diffuse emission plateau and
three major arms \citep{Angerhofer81, Mantovani85}. Its distance is
2~kpc \citep{Koo90}. A young energetic pulsar, PSR B1951+32, with $\rm
\dot{E} = 3.7\times10^{36}\ erg\ s^{-1}$ was discovered in the central
core region \citep{Kulkarni88}. Strong polarized emission of CTB~80
was observed, which varies with frequency \citep{Mantovani85}.
The overall radio spectral index is $\alpha = -0.45\pm0.03$
\citep{Kothes06b}. A possible spectral break at a low frequency below
1~GHz was noted by \citet{Mantovani85} and \citet{Castelletti03}.

The morphology of CTB~80 from the $\lambda$6\ cm observations (see
Fig.~\ref{SNRipi}) generally agrees with previous results. We see
faint extended emission from the eastern arm (E-arm) extending to
about $\ell \sim 70\fdg0$, in agreement with the observations by
\citet{Mantovani85}. This part of CTB~80 is highly polarized at
$\lambda$6\ cm, and seen in the polarization map at 1.4~GHz
\citep{Kothes06b}.  The apparent size of this SNR measured from the
$\lambda$6\ cm map is about $128\arcmin$ from east to west and about
$96\arcmin$ in the south-north direction, which implies that it has a
physical size of 74~pc $\times$ 56~pc at 2~kpc. In the $\lambda$6\ cm
map, we detected diffuse extended emission in the southwest between the
W-arm and S-arm.

We investigated the spectral index distribution in the central core
region and the three arms using Urumqi $\lambda$6\ cm and Effelsberg
$\lambda$21\ cm survey data for TT-plots (not shown in this paper). We
confirmed the flat spectrum in the central core with a brightness
temperature spectral index of $\beta = -1.97\pm0.20$.  We derived
$\beta = -2.49\pm0.03$ for the western arm (W-arm), $\beta =
-2.38\pm0.04$ for the southern arm (S-arm), and $\beta = -2.28\pm0.01$
for the eastern arm (E-arm) and the outer extension.

To get the spectrum of CTB~80 from integrated flux densities, we set
the SNR boundary according to the maps at four frequencies above 1~GHz
published by \citet{Mantovani85}, and outlined by the dashed line in
Fig.~\ref{SNRipi}. After subtracting all background radio sources, we
calculated an integrated flux density at $\lambda$6\ cm of $S_{\rm
  6cm} = 35.6\pm3.9$~Jy, slightly lower than that quoted by
\citet{Mantovani85}. In the same area, we got the $\lambda$11\ cm and
$\lambda$21\ cm flux densities from the Effelsberg survey data as
$S_{\rm 11cm} = 47.6\pm5.0$~Jy and $S_{\rm 21cm} = 56.7\pm6.3$~Jy,
respectively. Using these three well-determined new flux densities, we
got the spectral index of $\alpha = -0.38\pm0.13$ (see
Fig.~\ref{spectrum_SNR}). As noticed by \citet{Mantovani85}, the
spectral index of CTB~80 at higher frequencies should be considerably
flatter than previously thought \citep{Velusamy74, Sofue83}.

Flux densities available from literatures are shown in
Fig.~\ref{spectrum_SNR}. Taking the three newly obtained flux
densities into account, the spectrum of this SNR seems to be flat
below about 1~GHz. However, owing to the scattered data at low
frequencies the spectral shape is difficult to assess.
\citet{Castelletti05} discussed a possible low-frequency spectral
turn-over. However, free-free thermal absorption, which may produce
such a spectral break, would be inconsistent with the low $DM$ of the
central pulsar, PSR B1951+32 \citep[$\rm DM = 45~pc\ cm^{-3}$,
][]{Hobbs04}.

Intense $\lambda$6\ cm polarized emission pervades nearly the entire
SNR. We found B-field vectors running almost continuously along the
southern and the western arm, but vary in the eastern arm. The weaker
and discontinuous polarized emission in the middle of the eastern arm
is probably due to internal cancelation by the superposition of
polarization vectors orientated in different directions.  In the outer
extension of the eastern arm the percentage polarization is up to
50\%.  In the western arm, the polarization fraction is about 14\%,
while in the southern arm, the percentage increases from north to the
south with an average value of about 21\%.

\begin{figure}
\begin{center}
\includegraphics[angle=-90, width=0.4\textwidth]{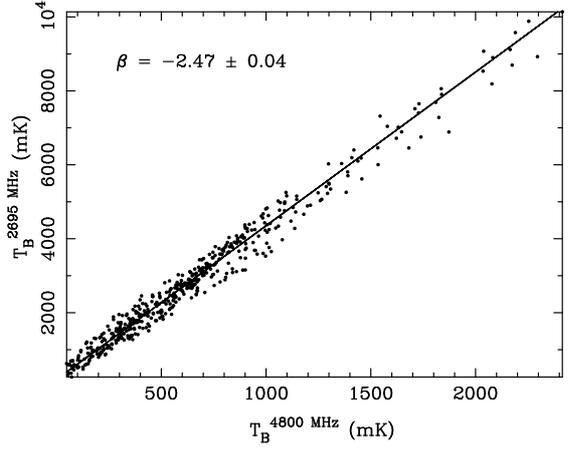}
\caption{TT-plot for G78.2+2.1 using the Urumqi $\lambda$6\ cm and
  Effelsberg $\lambda$11\ cm survey data.}
\label{G78_tt}
\end{center}
\end{figure}

\subsection{G78.2+2.1}

The SNR G78.2+2.1 is located in the direction of the Cygnus~X
region. It was suggested and confirmed to be a large SNR by
\citet{Higgs77a} and \citet{Higgs77b} based on radio maps at
$\lambda$3\ cm and $\lambda$21\ cm. Its distance is about 1.5~kpc
\citep{Landecker80}. A few \ion{H}{II} regions were detected near its
shell. The well-known \ion{H}{II} region $\gamma$ Cygni Nebula
overlaps with the bright southern shell. The \ion{H}{II} region
G78.3+2.8 \citep{Sabbadin76} is located to the north of the SNR, while
the \ion{H}{II} region IC 1318b is seen in the southeast direction.
Several methods have been tried to separate non-thermal and thermal
emission in the Cygnus~X area \citep{Wendker91, Zhang97,Ladouceur08}
to study the SNR emission without the dominating thermal background.
Spectral index variations within the SNR were discussed in detail by
\citet{Zhang97} and later by \citet{Ladouceur08} with a higher angular
resolution of $4\farcm5 \times 2\farcm9$.

The Urumqi $\lambda$6\ cm total intensity image (see
Fig.~\ref{SNRipi}) shows two bright unresolved shell sections in the
north and south and the central weak extended emission. Not many
detailed structures can be recognized because of the angular
resolution of $9\farcm5$. The boundary of the SNR is difficult to
determine in the Urumqi $\lambda$6\ cm map, since the shell sections
of the SNR merge with ambient \ion{H}{II} regions. We subtracted the
\ion{H}{II} region G78.3+2.8 from the map using Gaussian fitting, and
removed point-like sources as described in Sect.~3. After discounting
the flux density of the $\gamma$ Cygni Nebula, $S_{\rm 6cm}$ = 4.3~Jy
\citep{Wendker91}, we obtained an integrated flux density of this SNR
as $S_{\rm 6cm} = 170\pm18$~Jy, consistent with the value of $S_{\rm
  6cm} = 150\pm15$~Jy measured by \citet{Wendker91}. We derived $S_{\rm
  11cm} = 222\pm23$~Jy for the Effelsberg $\lambda$11\ cm survey
data. Combining the flux densities reported at other wavelengths, we
obtained an overall spectral index of $\alpha = -0.48\pm0.03$,
consistent with the result of \citet{Kothes06b}. The TT-plot for the
SNR using Urumqi $\lambda$6\ cm and Effelsberg $\lambda$11\ cm survey
data (Fig.~\ref{G78_tt}) gives $\beta=-2.47\pm0.04$, which agrees with
the spectrum derived from integrated flux intensities.
 
The $\lambda$6\ cm polarization data in this area are largely confused
with residual instrumental effects, e.g., some horizontal stripes from
the strong Cygnus~X region emission.

\begin{figure}
\begin{center}
\includegraphics[angle=-90, width=0.4\textwidth]{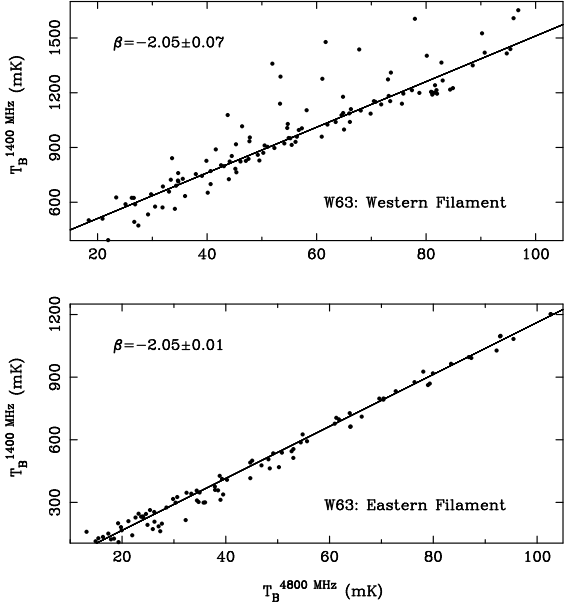}
\caption{TT-plot for the two filaments aside of G82.2+5.3 (W63)
  indicated in Fig.~\ref{SNRipi} using the Urumqi $\lambda$6\ cm and
  Effelsberg $\lambda$21\ cm data.}
\label{W63_tt}
\end{center}
\end{figure}

\subsection{G82.2+5.3 (W63)}

\begin{figure}
\begin{center}
\includegraphics[angle=-90, width=0.4\textwidth]{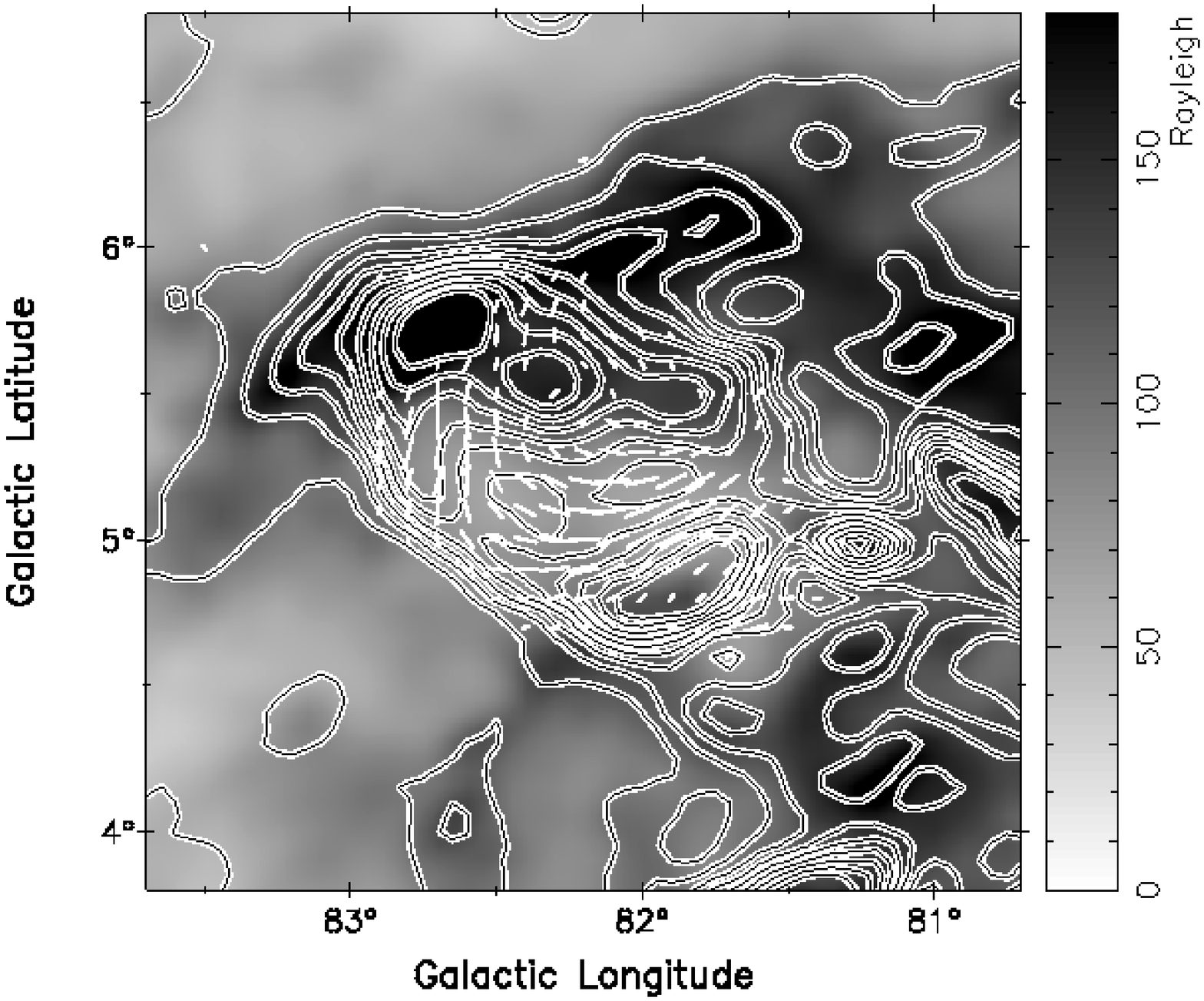}
\caption{H$\alpha$ emission \citep[from][]{Finkbeiner03} overlaid by
  Urumqi $\lambda$6\ cm total intensity contours and B vectors
  (i.e. $PA+90\degr$) from the SNR G82.2+5.3 (W63). Contours start at
  15.0~mK\ T$_{\rm B}$ and run in steps of 30.0~mK\ T$_{\rm
    B}$. B-field vector lengths are proportional to the polarization
  intensities.}
\label{w63_ha}
\end{center}
\end{figure}

W63 is a shell-type SNR located at a high Galactic latitude north of
the Cygnus~X complex. In this area, filamentary structures and strong
diffuse Galactic background emission exist, which might confuse with
emission from this SNR. Some radio filaments coincide with optical
emission. They are most probably thermal and unrelated to the SNR
\citep{Wendker71}. \citet{Angerhofer77} mapped W63 at $\lambda$6\ cm,
but did not present a complete map. They reported a lower flux density
limit of $S_{\rm 6cm} \geqslant 38.5\pm4.0$~Jy.  From the Urumqi
$\lambda$6\ cm observations, we obtained complete total intensity and
polarization maps for W63 at $\lambda$6\ cm (see
Fig.~\ref{SNRipi}). We checked the properties of the two major
filaments centered at $\ell = 81\fdg80, b =6\fdg10$ and $\ell =
83\fdg20, b =5\fdg60$ (marked with ``W-fila'' and ``E-fila'' in
Fig.~\ref{SNRipi}). The brightness-temperature spectral index from the
TT-plot (Fig.~\ref{W63_tt}) using the Urumqi $\lambda$6\ cm and
Effelsberg Medium Latitude Survey (EMLS) $\lambda$21\ cm data
\citep{Uyaniker99} is $\beta = -2.05\pm0.07$ for the western filament
(W-fila) and $\beta = -2.05\pm0.01$ for the eastern filament (E-fila)
(Fig.~\ref{W63_tt}). The H$\alpha$ map (Fig.~\ref{w63_ha}) clearly
shows the two bright filaments and their extensions beyond W63, which
proves their thermal nature. Therefore, we discounted these two
filaments in the flux density integration and got $S_{\rm 6cm} =
48.5\pm4.9$~Jy. We also calculated the integrated flux density at
$\lambda$21\ cm for the same area from the EMLS \citep{Uyaniker99},
$S_{\rm 21cm} = 93.1\pm9.5$~Jy. Combining all available flux
densities, we obtained a spectral index of $\alpha = -0.44\pm0.04$.

Strong polarized $\lambda$6\ cm emission was detected within the SNR,
where the strongest polarized emission coincides with the area of
minimum total intensity, similar to what was seen at $\lambda$11\ cm
by \citet{Velusamy74}. Polarized emission at $\lambda$21\ cm is
visible within the SNR with a percentage polarization of about 4\% to
11\% in the area, where the $\lambda$6\ cm polarization is up to 66\%.
We noticed that the H$\alpha$ emission \citep{Finkbeiner03} is
relatively low in an area almost coinciding with the high polarization
of W63 (see Fig.~\ref{w63_ha}). This suggests that the observed
H$\alpha$ emission is probably in the foreground of W63 and causes
depolarization in other parts of the SNR.

\subsection{G89.0+4.7 (HB21)}

HB21 is a strong, large, and evolved SNR with an unusual complex
morphology (see Fig.~\ref{SNRipi}). It has been observed at many
frequencies from 38~MHz \citep{Crowther65} to 5~GHz
\citep{Kundu71}. The distance to the SNR was estimated to be about
$800\pm70$~pc \citep{Tatematsu90}. HB21 resides in an area of
relatively low Galactic emission and has a well-defined outer
boundary, which allows us to obtain integrated flux densities without
any confusion problem.

HB21 was partly covered by the 1.4~GHz polarization survey of the
Galactic plane by \citet{Landecker10} with $1\arcmin$ angular
resolution, where data from the DRAO interferometer were combined with
Effelsberg 100-m and DRAO 26-m telescope data. At that frequency, HB21
appears almost depolarized with the possible exception of a few
discrete patches in this direction.

The $\lambda$6\ cm total intensity and polarization intensity maps
resemble the previous $\lambda$6\ cm observations presented by
\citet{Kundu71}. We obtained an integrated flux density of
$\lambda$6\ cm map as $S_{\rm 6cm} = 107\pm11$~Jy. Including other
available flux densities for a wide frequency range (see references in
Table~1), we fitted a spectral index of $\alpha = -0.36\pm0.03$.  The
$\lambda$6\ cm polarization map shows that polarization angles are not
regularly ordered. Some depolarized `canals' are visible, where the
$PA$s on both sides differ by $90\degr$.  At the angular resolution of
$9\farcm5$ of the $\lambda$6\ cm observations, details are smoothed
out.

\subsection{G93.7$-$0.2}

SNR G93.7$-$0.2 is a shell-type SNR with two major thick brightened
limbs and a few diffuse extensions. The extensions were regarded as
possible ``break-out'' phenomenon \citep{Uyaniker02}.  Polarized
emission of G93.7$-$0.2 is found to be widely spread at 4.75~GHz
\citep{Mantovani91} and patchy at 1.4~GHz because of depolarization
\citep{Uyaniker02}.  The distance to G93.7$-$0.2 was found to be about
1.5~kpc \citep{Uyaniker02}.

The $\lambda$6\ cm $I$ map (Fig.~\ref{SNRipi}) shows all extensions
previously reported. The eastern extension at about $\ell = 94\fdg50,
b = 0\fdg30$ is not part of the SNR but is an \ion{H}{II} region
\citep{Uyaniker02}. We found another \ion{H}{II} region at $\ell =
94\fdg85, b = -0\fdg05$, where the TT-plot between the Urumqi
$\lambda$6\ cm and the Effelsberg $\lambda$21\ cm survey data (not
shown in this paper) gave a brightness-temperature spectral index of
$\beta = -2.11\pm0.11$. Discarding point-like sources and the thermal
extensions, we obtained an integrated flux density of $S_{\rm 6cm} =
25.0\pm2.5$~Jy, lower than that reported by \citet{Mantovani91}. We
obtained $S_{\rm 21cm} = 39.6\pm4.0$~Jy and $S_{\rm 11cm} =
35.9\pm3.6$~Jy from the Effelsberg survey data in the same
area. Combining these new flux densities with those in the literature,
we obtained an overall spectral index for G93.7$-$0.2 of $\alpha =
-0.52\pm0.03$.

The polarized emission of SNR G93.7$-$0.2 we detected at
$\lambda$6\ cm agrees with the observational results by
\citet{Mantovani91}.

\subsection{G106.3+2.7}

G106.3+2.7 is a comet-shaped SNR discovered by \citet{Joncas90} from
the DRAO northern Galactic plane survey at 408~MHz, and further
studied by \citet{Pineault00}. The SNR consists of a compact ``head''
region containing an off-center PWN G106.6+2.9 in the north, the
so-called ``Boomerang'' and a diffuse ``tail'' region (see
Fig.~\ref{SNRipi}).  On the basis of the linearly polarized emission
of the SNR, \ion{H}{I}, and CO measurements, \citet{Kothes01} argued
that both the PWN and the SNR resulted from the same SN event at a
distance of 800~pc. The ``head'' region is created by the interaction
between the expanding blast wave and the ambient dense material, while
the ``tail'' region is an outbreak to the interior of an \ion{H}{I}
bubble.  \citet{Kothes06a} performed a multi-frequency analysis
towards the PWN and found a spectral break at about 4.3~GHz. The
overall spectral index of the whole SNR is $\alpha \sim -0.6$, while
the compact ``head'' region has a flatter spectrum than the ``tail''
region \citep{Pineault00}.

Observations of the entire SNR at $\lambda$6\ cm have never been made.
The Urumqi $\lambda$6\ cm map shows a compact ``head'', while the
elongated ``tail'' region is confused by diffuse emission in the
west. The PWN, as indicated by a star in Fig.~\ref{SNRipi}, cannot be
resolved in the $\lambda$6\ cm map. To determine the integrated flux
density of G106.3+2.7 at $\lambda$6\ cm, we adapt the integration
boundary from the Effelsberg $\lambda$21\ cm survey map, which can be
more clearly defined.  We obtained $S_{\rm 6cm} = 2.0\pm0.3$~Jy from
the Urumqi map, and both $S_{\rm 11cm} = 3.1\pm0.3$~Jy and $S_{\rm
  21cm} = 5.0\pm0.6$~Jy from the Effelsberg maps. Using these flux
densities, together with published flux densities at other
frequencies, we obtained an integrated spectral index of $\alpha =
-0.64\pm0.04$.

Strong polarized emission of SNR G106.3+2.7 was detected at
$\lambda$6\ cm. The average percentage polarization is about 25\% with
maxima around 42\%.

\begin{figure}
\begin{center}
\includegraphics[angle=-90, width=0.4\textwidth]{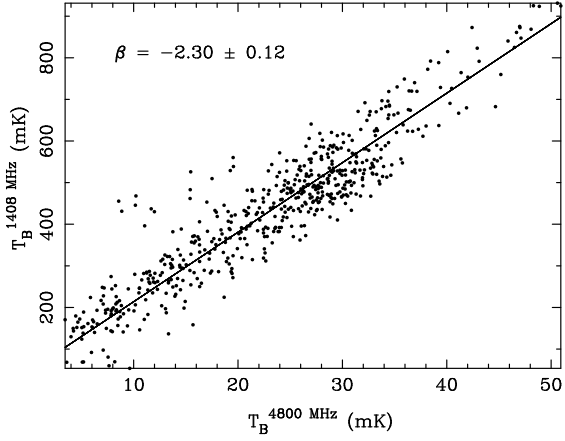}
\caption{TT-plot of the SNR G114.3+0.3 using the Urumqi $\lambda$6\ cm
  and Effelsberg $\lambda$21\ cm survey data.}
\label{G114_tt}
\end{center}
\end{figure}

\subsection{G114.3+0.3}

G114.3+0.3 was identified as a low surface brightness shell-type SNR
by \citet{Reich81}.  It is located in the local arm at about 700~pc
distance \citep{Yar-Uyaniker04}. The pulsar B2334+61, indicated by a
star in Fig.~\ref{SNRipi}, is located near the SNR center, and
physically associated with the SNR \citep{Fuerst93,Kulkarni93}.  The
spectral index of this SNR cannot be clearly determined because of its
low surface brightness.  \citet{Reich81} quoted a spectral index
$\alpha = -0.31\pm0.1$ from flux densities at $\lambda$21\ cm and
$\lambda$11\ cm. \citet{Tian06y} used the TT-plot method for the
408~MHz and 1420~MHz CGPS data and obtained $\alpha = -0.16\pm0.41$.

No other $\lambda$6\ cm observations of G114.3+0.3 have so far been
reported. The map of this elliptical-shaped SNR is presented in
Fig.~\ref{SNRipi}. The ridges of strong continuum emission in the
southwest and southeast directions are clearly visible.  After
removing all point-like sources including the bright \ion{H}{II}
region, SH 2-165, in the eastern part of SNR \citep{Reich81}, we
calculated an integrated flux density of $S_{\rm 6cm} =
6.9\pm0.7$~Jy. From the Effelsberg $\lambda$21\ cm survey, we obtained
an integrated flux density $S_{\rm 21cm} = 10.8\pm1.1$~Jy, which
agrees with that obtained by \citet{Tian06y}. Using these flux density
data, as well as the value measured at 408~MHz from \citet{Tian06y}
despite its large uncertainties, we got a spectral index of $\alpha =
-0.31\pm0.10$ (see Fig.~\ref{spectrum_SNR}). This spectrum index can
be verified by the TT-plot between the Urumqi $\lambda$6\ cm and
Effelsberg $\lambda$21\ cm survey data as $\beta = -2.30\pm0.12$
(Fig.~\ref{G114_tt}).

Polarized emission of this SNR mainly comes from the central and
southern parts, and is very strong along the southern edge.  When we
compared the $PA = -15\degr$ at $\lambda$11\ cm \citep{Reich81} in a
region centered at about $\ell = 114\fdg50, b = 0\fdg25$ with the $PA =
45\degr$ we observed at $\lambda$6\ cm, we found that $RM =
-120$~rad~m$^{-2}$ for the SNR, which is consistent with $RM$ of the
pulsar PSR B2334+61, $RM = -100\pm18~$rad\ m$^{-2}$ \citep{Mitra03}.

\begin{figure}
\begin{center}
\includegraphics[angle=-90, width=0.4\textwidth]{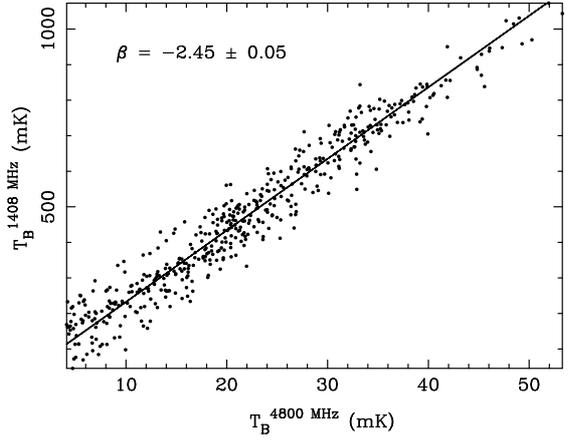}
\caption{The same as shown in Fig.~\ref{G114_tt}, but for the SNR
  G116.5+1.1.}
\label{G116_tt}
\end{center}
\end{figure}

\subsection{G116.5+1.1}

G116.5+1.1 is a shell-type SNR discovered by \citet{Reich81}. Its
distance is 1.6~kpc \citep{Yar-Uyaniker04}. \citet{Tian06y} derived a
spectral index of $\alpha = -0.28\pm0.15$ from the 408~MHz and
1420~MHz CGPS data, while \citet{Kothes06b} calculated $\alpha =
-0.16\pm0.11$. By supplementing the flux densities measured at
$\lambda$21\ cm and $\lambda$11\ cm \citep{Reich81}, \citet{Kothes06b}
found a steeper overall spectral index of $\alpha
=-0.53\pm0.08$.

We obtained the first total intensity and polarization measurements
for G116.5+1.1 at $\lambda$6\ cm. The integrated flux density was
found to be $S_{\rm 6cm} = 5.7\pm0.6$~Jy, We also obtained $S_{\rm
  21cm} = 10.3\pm1.1$~Jy from the Effelsberg $\lambda$21\ cm survey
data and calculated an overall radio spectral index of $\alpha =
-0.45\pm0.05$. The TT-plot was used to verify this result (see
Fig.~\ref{G116_tt}).

Intense polarized emission is detected from the western and northern
parts of the shell of this SNR. In the western part, B-field vectors
(i.e. $PA$+$90\degr$) are aligned tangentially to the shell at
$\lambda$6\ cm. By comparing them with the $PA$s observed at
$\lambda$11\ cm by \citet{Reich81} (see their Fig.~2b), we estimated
the $RM$. The $PA$ values in the region around $\ell = 116\fdg05, b =
1\fdg25$ at $\lambda$11\ cm and $\lambda$6\ cm are $PA_{\rm 11cm} \sim
40\degr$ and $PA_{\rm 6cm} \sim 100\degr$, so the calculated $RM \sim
-120\ $rad~m$^{-2}$, is the same as that for SNR G114.3+0.3.

\subsection{G160.9+2.6 (HB9)}

HB9 is a strong large nearby SNR that has been mapped at radio
frequencies from 10~MHz \citep{Caswell76} to 5~GHz
\citep{DeNoyer74}. We note that the designation G160.9+2.6 is based on
early radio maps. The center of HB9 is at $\ell = 160\fdg4, b =
2\fdg8$. The SNR has a distance of $0.8\pm0.4$~kpc
\citep{Leahy07}. HB9 has multiple fragmented shells and internal
filamentary structures. \citet{Reich03} reported a spectral index of
$\alpha = -0.57\pm0.03$ based on a TT-plot between 865~MHz data and
4750~MHz Effelsberg data. \citet{Kothes06b} compiled integrated flux
densities at many frequencies and obtained a spectral index of $\alpha
= -0.64\pm0.02$.

From the Urumqi $\lambda$6\ cm survey we calculated an integrated flux
density of $S_{\rm 6cm} = 33.9\pm3.4$~Jy, which is consistent with the
value of 36$\pm$8~Jy previously reported by \citet{DeNoyer74} at
5~GHz.  Using our $\lambda$6\ cm flux density and those at lower
frequencies (see Table~1), we calculated a spectral index of $\alpha =
-0.59\pm0.02$, in agreement with that of \citet{Reich03}.

The polarized emission of HB9 at $\lambda$6\ cm was previously mapped
with the 100-m Effelsberg telescope to find the intrinsic magnetic
field configuration within the SNR, by using in addition polarization
data at $\lambda$11\ cm and $\lambda$21\ cm \citep{Fuerst04}. Our
polarization map at $\lambda$6\ cm of HB9 has lower angular
resolution. Strong polarization is seen in a few patches near the
southern, northern, and eastern periphery of the shell.

\subsection{G166.0+4.3 (VRO 42.05.01)}

VRO 42.05.01 has an unusual morphology as it consists of two shells
with significantly different radii (see Fig.~\ref{SNRipi}).
\citet{Pineault87} interpreted this unusual morphology by modeling the
southern shell (wing region) expanding from the dense warm medium
surrounding the northern shell into a hot tunnel of very low density
(see their Fig.~3). \citet{Kothes06b} calculated an overall radio
spectral index of $\alpha = -0.37\pm0.03$. \citet{Leahy05} studied the
distribution of the spectral indices across the SNR and found that the
spectral index in the southern wing is steeper than elsewhere.

We measured an integrated flux density of $S_{\rm 6cm} =
3.3\pm0.3$~Jy, and derived an overall spectral index of $\alpha =
-0.33\pm0.04$ using available data (see references in Table~1), which
is consistent with previous results. By comparing the Urumqi
$\lambda$6\ cm and Effelsberg $\lambda$21\ cm survey data, we confirm
that the southern wing has a steeper spectral index. We calculated a
brightness-temperature spectral index of $\beta_{\rm 6-21} =
-2.33\pm0.05$ for the northern shell (head region) and $\beta_{\rm
  6-21} = -2.49\pm0.09$ for the southern wing.

Both the northern head region and the southern wing region of VRO
42.05.01 emit strong polarized emission. B-field vectors in the head
region and the eastern part of the wing region run almost tangential
to the shell, while significant deviations are seen for the western
part of the wing region indicating a significant $RM$ within the
SNR. Assuming the initial B-fields run tangential along the wing, the
deviation of B-field vectors indicates that $|RM|$ is about
130~rad\ m$^{-2}$.

\subsection{G179.0+2.6}

G179.0+2.6 is a thick shell-type SNR identified and studied by
\citet{Fuerst86} using $\lambda$21\ cm, $\lambda$11\ cm, and
$\lambda$6\ cm data observed with the Effelsberg 100-m telescope. A
polarized triple source near the center of the SNR was identified from
high angular resolution VLA observations and by optical identification
as a background double-sided radio galaxy \citep{Fuerst89}. The
overall spectral index of the SNR was found to be about $\alpha =
-0.30\pm0.15$.

After we discounted the flux contribution of 15 point-like sources
listed in the NVSS catalogue \citep{Condon98}, we obtained an
integrated flux density of SNR G179.0+2.6 as $S_{\rm 6cm} =
3.2\pm0.3$~Jy. The flux densities we obtained from the Effelsberg
$\lambda$21\ cm and $\lambda$11\ cm survey are $S_{\rm 21cm} =
5.4\pm0.6$~Jy and $S_{\rm 11cm} = 5.0\pm0.5$~Jy. Because
\citet{Fuerst86} removed only the strongest source 4C 31.21 within the
area of the SNR when calculating the integrated flux density, our
results with an improved assessment of background source contribution
are more accurate and reveal an overall spectral index for SNR
G179.0+2.6 of $\alpha = -0.45\pm0.11$ (see Fig.~\ref{spectrum_SNR}).

A radial configuration of the magnetic field of SNR G179.0+2.6 was
revealed by \citet{Fuerst86} that is confirmed by the Urumqi
$\lambda$6\ cm polarization observations. Such a magnetic field
configuration is usually seen in young SNRs, although the low surface
brightness of G179.0+2.6 indicates that it is an evolved object. When
the SNR is barrel-shaped, but viewed along its poles as described by
\citet{Whiteoak68} the magnetic field appears radial, although the
magnetic field lines are tangential to the compressed shell. The
observed radio emission is weak. This scenario would require that the
regular Galactic magnetic field in the area of G179.0-2.6 is
orientated along the line of sight. This is unexpected for the
Galactic anti-center, where the regular magnetic field is believed to
run almost tangential to the line of sight.

\subsection{G189.1+3.0 (IC443)}

IC443 is a ``mixed-morphology'' type SNR \citep{Rho98}, that is
distinct from the three well-defined SNR types: shell-like, Crab-like,
and composite. It interacts with a molecular cloud \citep{Burton88,
  VDishoeck93, Cesarsky99}. This SNR has been extensively observed
throughout the whole electromagnetic spectrum \citep[e.g.][]{Mufson86,
  Albert07}. \citet{Kundu68} collected flux densities at five
frequencies from 430~MHz to 5~GHz, and suggested a spectral break
around 750~MHz towards the brightened northwestern shell.
\citet{Erickson85} proposed a possible overall spectral break below
20~MHz. According to \citet{Green86b}, spectral indices vary across
IC443. \citet{Leahy04} found that the brightened northwestern shell
region has a radio spectral index of $\alpha = -0.43\pm0.02$, while in
the fainter southern part $\alpha$ varies with values ranging from
$-0.2$ to $-0.6$.

As in the $\lambda$11\ cm map by \citet{Fuerst90}, the $\lambda$6\ cm
map of IC443 in Fig.~\ref{SNRipi} shows strong radio emission from the
SNR, as outlined by the dashed line, and also the weaker unpolarized
thermal emission from the extended \ion{H}{II} region SH 2-249 north
of the remnant. For IC443, we obtained an integrated flux density from
the $\lambda$6\ cm map of $S_{\rm 6cm} = 84.6\pm9.4$~Jy. Together with
flux densities at other frequencies (see references in Table~1), we
calculate an overall radio spectral index of $\alpha =-0.38\pm0.01$.

The Urumqi $\lambda$6\ cm polarization image (Fig.~\ref{SNRipi}) shows
a similar dipole-shaped magnetic field configuration as earlier noted
by \citet{Kundu69} and \citet{Dickel76}. The radial configuration of
the magnetic fields observed at $\lambda$6\ cm seems to be
intrinsic. The question is whether and how the field structure is
related to the PWN inside IC443. \citet{Olbert01} and
\citet{Bocchino01} identified the PWN from hard X-ray observations
with a compact head at about $\ell = 189\fdg2, b = 2\fdg9$ near the
center of the dipole-shaped $B$ fields (as indicated by a star in
Fig.~\ref{SNRipi}).  \citet{Gaensler06} found that inside IC 443 the
X-ray source G189.22+2.9 is a thermally emitting neutron star moving
through the hot plasma. The distance to the SNR IC443 was noted to be
about 1.5~kpc \citep{Fesen84}. At this distance, a $1\degr$ scale
corresponds to the size of 26\ pc, much larger than the typical PWN
size of a few pc. It is thus unlikely that the PWN is responsible for
the whole dipole-shaped B field observed (see
Fig.~\ref{SNRipi}). However, our data have insufficient angular
resolution to clarify the role of the PWN inside IC443 and determine
the origin of the radial magnetic field configuration seen in IC443.

From the X-ray image of ROSAT, \citet{Asaoka94} detected a large
($\sim 1.5\degr$ in diameter) and very faint X-ray shell that
overlapped with IC~443. They proposed that this shell is another SNR,
G189.6+3.3. Near the IC443 boundary, a filament was seen both in the
optical image from the Digital Sky Survey and the radio at 1.4~GHz
\citep{Leahy04} that might be a part of this proposed SNR. At
$\lambda$6\ cm, only a very short narrow radio rim appears at $l =
189\fdg1, b = 3\fdg6$, i.e. along the optical filament emerging from
IC~443. This short rim is also seen in the Effelsberg $\lambda$11\ cm
and $\lambda$21\ cm surveys. Using the TT-plot, we obtained
$\beta_{\rm 6-11} = -2.49\pm0.25$ and $\beta_{\rm 6-21} =
-2.35\pm0.22$. No radio emission from other parts of the proposed SNR
G189.6+3.3 is detected at $\lambda$6\ cm and the Effelsberg
surveys. Therefore, these observations cannot clarify the existence of
the proposed {\it large} SNR G189.6+3.3.

\begin{figure}
\begin{center}
\includegraphics[angle=-90, width=0.4\textwidth]{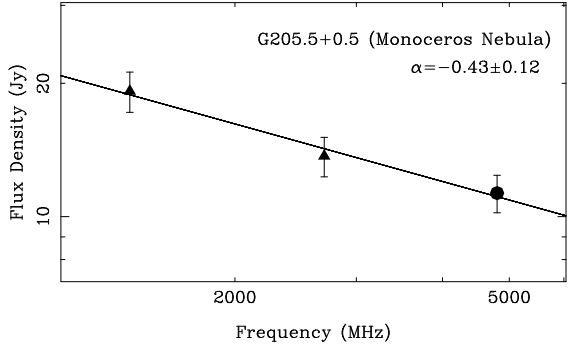}
\caption{Integrated radio flux densities of the western shell of SNR
  G205.5+0.5 (Monoceros Nebula) (as marked in Fig.~\ref{SNRipi2}).}
\label{Mon_spectrum}
\end{center}
\end{figure}

\subsection{G205.5+0.5 (Monoceros Nebula)} 

The SNR Monoceros Nebula is located in the area of the bright Rosette
Nebula (SH 2-275), a famous \ion{H}{II} region. \citet{Odegard86}
suggested that the SNR is at the same distance as the \ion{H}{II}
region of about 1.6~kpc and that both objects interact with each
other. On the basis of the radio absorption, he also argued that the
SNR is situated behind the \ion{H}{II} region. The SNR was studied at
low and intermediate radio frequencies by \citet{Holden68},
\citet{Milne69}, and \citet{Dickel75}. The radio structures are quite
complicated in this region. The Rosette Nebula overlaps with the
southern part of the SNR. The southern ridge of the \ion{H}{II} region
SH 2-273 is also mixed with the edge of the western shell of the SNR
as shown by \citet{Graham82}. The extended nebula 0641+06 with the
center at about $\ell = 206\fdg35, b = 1\fdg35$ was identified to be
an \ion{H}{II} region \citep{Graham82} and be unrelated to the
SNR. Therefore, it was difficult to determine the boundary of the SNR,
and reliable integrated flux densities could not be obtained. Early
$\lambda$11\ cm polarization observations were made by
\citet{Milne74}, but no significant polarized emission was detected.

The $\lambda$6\ cm Urumqi survey provides the shortest wavelength
observations of the Monoceros Nebula and shows polarized emission,
mainly towards the western shell of the SNR. After subtracting
point-like sources, we obtained flux densities of the shell
(integration area indicated in Fig.~\ref{SNRipi2}) from the Urumqi
$\lambda$6\ cm, Effelsberg $\lambda$11\ cm, and $\lambda$21\ cm survey
data. These were well fitted by a non-thermal spectrum with $\alpha =
-0.43\pm0.12$ (see Fig.~\ref{Mon_spectrum}).

\subsection{G206.9+2.3 (PKS 0646+06)}

G206.9+2.3 is a discrete SNR close to the Monoceros Nebula that has a
bright northwestern radio shell. The distance was estimated to be from
about 3~kpc to 5~kpc \citep{Graham82}. In earlier studies, G206.9+2.3
was believed to be a possible extension of the Monoceros Nebula, until
\citet{Davies78} identified it as a distinct SNR based on optical
observations. \citet{Graham82} quoted flux densities at different
radio frequencies from 38~MHz to 2.7~GHz and obtained a spectral index
of $\alpha = -0.45\pm0.03$.  On the basis of this result,
\citet{Odegard86} argued for a spectral turnover below 38~MHz because
of the lower flux density he measured at 20.6~MHz.

Observations of this SNR at $\lambda$6\ cm or shorter wavelength bands
are not yet available. The Urumqi survey showed strong polarized
emission of up to 27\% in the northwestern shell of the SNR (see
Fig.~\ref{SNRipi2}). The B-field vectors are not tangential to the
shell. The $RM$ value for the shell of PKS 0646+06 may be high if the
intrinsic B-field direction follows the shell.

After subtracting point-like sources including the strong point source
NVSS J064935+053823 in the east of the SNR G206.9+2.3, we obtained a
flux density of $S_{\rm 6cm} = 2.9\pm0.3$~Jy from the Urumqi
$\lambda$6\ cm, and $S_{\rm 11cm} = 3.9\pm0.4$ and $S_{\rm 21cm} =
5.1\pm0.6$~Jy from the Effelsberg $\lambda$11\ cm and $\lambda$21\ cm
survey data, respectively. Combining the available flux densities in
the literature, we got a spectral index of $\alpha = -0.47\pm0.04$,
which agrees with that derived by \citet{Graham82}.

\section{Summary}

We have studied 16 SNRs larger than $1\degr$ based on maps from the
Sino-German $\lambda$6\ cm polarization survey of the Galactic plane
carried out with the Urumqi 25-m telescope. We have found that
G192.8$-$1.1 consists of compact sources and thermal components, and
conclude that G192.8$-$1.1 is not a SNR.
 
For the SNRs, we obtained integrated flux densities when possible,
derived their spectral indices, and analyzed the polarization data.
Our results for the SNRs W50, G78.2+2.1, HB21, and IC443, are
consistent with previous observations.  The Urumqi $\lambda$6\ cm flux
densities of SNRs, G65.1+0.6, W63, G93.7$-$0.2, G106.3+2.7,
G114.3+0.3, G116.5+1.1, HB9, VRO~42.05.01, and G206.9+2.3, are those
of the highest frequency available and are valuable when interpreting
the SNR spectra.  All SNRs are polarized by more than 10\% with the
exception of G78.2+2.1, which is located towards the thermal Cygnus
region and IC443.  The polarization images of W63, G106.3+2.7,
G114.3+0.3, G116.5+1.1, HB9, and VRO~42.05.01, are those observed at
the highest frequency so far, including those of the Monoceros Nebula
and G206.9+2.3, which are the first polarization images.

At $\lambda$6\ cm, the polarization observations trace the intrinsic
magnetic field configuration of SNRs when Faraday rotation can be
neglected, which seems to be the case for most SNRs.

The $\lambda$6\ cm total intensity and polarization data of the 16
SNRs and G192.8$-$1.1 are available from the NAOC
website\footnote{http://zmtt.bao.ac.cn/6cm/}.

\begin{acknowledgements}
We like to thank the anonymous referee for constructive comments. XYG
thanks the joint doctoral training plan between CAS and MPG and
financial support from CAS and MPIfR. The Sino-German $\lambda$6\ cm
polarization survey was carried out with a receiver system constructed
by Mr. Otmar Lochner at MPIfR mounted at the Nanshan 25-m telescope of
the Urumqi Observatory of NAOC. The MPG and the NAOC/CAS supported the
construction of the receiving system by special funds. We thank
Mr. Maozheng Chen and Mr. Jun Ma for qualified maintenance of the
receiving system for many years. The Chinese authors are supported by
the National Natural Science foundation of China (10773016, 10821061,
and 10833003), the National Key Basic Research Science Foundation of
China (2007CB815403) and the Partner group of the MPIfR at NAOC in the
frame of the exchange program between MPG and CAS for many bilateral
visits. XHS acknowledges financial support by the MPG and by
Prof. M. Kramer during his stay at MPIfR.
\end{acknowledgements}

\bibliographystyle{aa}
\bibliography{bbfile}

\end{document}